\documentclass[aps,pra,superscriptaddress,showpacs,twocolumn,10pt]{revtex4-1}
\usepackage{bm}
\usepackage{amsmath}
\usepackage{textcomp}
\allowdisplaybreaks
\usepackage{longtable}
\usepackage{dcolumn}
\usepackage{nicefrac}
\usepackage{graphicx}
\usepackage{xcolor}

\newcolumntype{.}{D{x}{}{-1}}

\newcolumntype{w}[1]{D{.}{.}{#1}}

\newcommand{\balpha}{\bm{\alpha}}
\newcommand{\bnabla}{\bm{\nabla}}

\newcommand{\bsigma}{\bm{\sigma}}
\newcommand{\bfx}{\bm{x}}

\newcommand{\bfr}{\bm{r}}

\newcommand{\Za}{{Z\alpha}}

\newcommand{\vare}{\varepsilon}

\newcommand{\lbr}{\langle}
\newcommand{\rbr}{\rangle}

\newcommand{\Dmatrix}[4]{
        \left(
        \begin{array}{cc}
        #1  & #2   \\
        #3  & #4   \\
        \end{array}
        \right)
        }

\newcommand{\SixJ}[6]{
        \left\{
        \begin{array}{ccc}
        #1  & #2  & #3 \\
        #4  & #5  & #6 \\
        \end{array}
        \right\}
        }

\newcommand{\I}[4]{I_{#1 #2 #3 #4}}
\newcommand{\W}[4]{W_{#1 #2 #3 #4}}
\newcommand{\U}[2]{U_{#1 #2}}

\newcommand{\intinf}{\int^{\infty}_{-\infty}}

\begin{document}

\title{QED calculation of the $\bm{2p}$ fine structure in Li-like ions}

\author{Vladimir A. Yerokhin}
\affiliation{Center for Advanced Studies, Peter the Great St.~Petersburg Polytechnic University,
Polytekhnicheskaya 29, 195251 St.~Petersburg, Russia}

\author{Mariusz Puchalski}
\affiliation{Faculty of Chemistry, Adam Mickiewicz University, Umultowska 89b, 61-614 Pozna{\'n},
Poland}

\author{Krzysztof Pachucki}
\affiliation{Faculty of Physics, University of Warsaw,
             Pasteura 5, 02-093 Warsaw, Poland}

\begin{abstract}

Large-scale {\em ab initio} QED calculations are performed for the $2p_{3/2}$--$2p_{1/2}$
fine-structure interval of Li-like ions with nuclear charges $Z = 5\,$--$\,92$. Improved
theoretical predictions are obtained by combining together two complementary theoretical
methods, namely, the approach that accounts for all orders in the binding nuclear strength and
the nonrelativistic QED approach that accounts for all orders in the nonrelativistic
electron-electron interaction. The resulting unified approach provides  theoretical predictions
which are more accurate than the available experimental results across the interval of the
nuclear charges considered.

\end{abstract}

\maketitle

Three-electron atoms, namely, Li and Li-like ions, are among the simplest many-electron systems.
They can be calculated {\em ab initio} within quantum electrodynamics (QED)  and measured with a
very high precision. Investigations of such atoms enable precision tests of bound-state QED of
many-body systems and allow studies of nuclear properties probed by atomic electrons
\cite{indelicato:19}. The spectacular experimental progress achieved during the past decades in
spectroscopy of Li-like atoms
\cite{schweppe:91,beiersdorfer:98,feili:00,brandau:03,beiersdorfer:05,epp:07,lestinsky:08,zhang:08,brown:13,nortenhauser:15}
motivated large efforts devoted to QED calculations of energy levels in these systems.

There are presently two main {\em ab initio} methods that systematically describe various atomic
properties within QED. The first method, described in Ref.~\cite{shabaev:02:rep}, accounts for
all orders in the nuclear binding strength (i.e., the parameter $\Za$, where $Z$ is the nuclear
charge number and $\alpha$ is the fine-structure constant) but expands in the number of virtual
photons exchanged between the electrons (i.e., in the parameter $1/Z$). Such calculations were
performed by a number of authors, most notably, by the Notre-Dame
\cite{blundell:93:a,sapirstein:01:lamb,sapirstein:11} and the St.~Petersburg
\cite{yerokhin:99:sescr,artemyev:99,yerokhin:00:prl,yerokhin:01:2ph,artemyev:03,yerokhin:07:lilike,kozhedub:10}
group. This method yields very accurate results for high-$Z$ ions, providing one of the best
tests of QED in the strong-field regime \cite{yerokhin:06:prl}. In the low-$Z$ region, however,
the applicability of this method diminishes, since the relative contribution of the electron
correlation increases as $Z$ goes down and the convergence of the $1/Z$ expansion deteriorates.

For light atoms, the  best results are  obtained with the second method, based on the
nonrelativistic quantum electrodynamics (NRQED) \cite{caswell:86}. This method expands the energy
levels of a bound system in powers of $\alpha$ and $Z\alpha$, but accounts for all orders in
$1/Z$. High-precision NRQED calculations were performed for energy levels of Li and Be$^+$ in
Refs.~\cite{yan:95:li,yan:98:prl,yan:02,puchalski:08:li,puchalski:09,puchalski:14}. For heavier
systems, however, the accuracy of the NRQED results deteriorates as $Z$ increases, since the
omitted higher-order effects become enhanced by high powers of $Z$.

The fine structure (fs) of energy levels is particularly favourable for theoretical calculations
by the NRQED method, offering numerous simplifications. For example, only a few operators
explicitly depending on the electron spin contribute to the fs splitting at the leading order of
the NRQED expansion, $m\alpha^4$ (where $m$ is the electron mass). Furthermore, at the
next-to-leading order $m\alpha^5$, the leading QED contribution comes only from the anomalous
magnetic moment of the electron. Owing to these and other theoretical simplifications, the $2p$
fs interval in Li and Be$^+$ is presently calculated up to order $m\alpha^6$
\cite{puchalski:14,puchalski:15}, while for other energy intervals of three-electron systems the
$m\alpha^6$ effects remain uncalculated so far.

In the present investigation we will combine the $1/Z$-expansion method and the NRQED approach
and obtain the most accurate theoretical predictions for the $2p_{3/2}\,$--$\,2p_{1/2}$ fs
interval through the lithium isoelectronic atomic sequence with $Z \ge 5$. To this end, we will
match the $\Za$ expansion of numerical results obtained by the $1/Z$-expansion method and the
$1/Z$ expansion of the NRQED results. The main improvement will be achieved in the region of
medium nuclear charges, $Z$$\,\approx\,$$8\,$--$\,20$, in which the both above-mentioned methods
do not work well.

The relativistic units ($\hbar=c=m=1$) will be used throughout this paper, unless explicitly
specified otherwise.

\section{$\bm{1/Z}$-expansion QED}
\label{sec:allorder}

In the present work, theoretical contributions to the energy of a Li-like atom are separated into
three parts, namely, the electron-structure part $E_{\rm struc}$, the radiative QED correction
$E_{\rm rad}$, and the nuclear recoil correction $E_{\rm rec}$,
\begin{align}\label{eq:1}
E = E_{\rm struc} +  E_{\rm rad} +  E_{\rm rec}\,.
\end{align}
We note that we distinguish between the QED effects of the self-energy and vacuum-polarization
type (termed as the radiative QED effects, $E_{\rm rad}$) and the QED effects originating from
the frequency-dependence of the electron-electron interaction (termed as the electron-structure
QED effects and included into $E_{\rm struc}$).

The $2p$ fs splitting of Li-like atoms is obtained as a difference of energies of the $2p$ states
with different values of the total angular momentum, $(1s)^22p_{3/2}$ and $(1s)^22p_{1/2}$. In
the following, we will denote by $E_i(v)$ corrections to the ionization energy of the valence
electron state $v$ and by $E_i({\rm fs})$ corrections to the fs splitting, $E_i({\rm fs}) =
E_i(2p_{3/2}) - E_i(2p_{1/2})$. We note that the energy contributions involving interactions only
between the core electrons do not contribute neither to the ionization energy or the fs interval,
so they are not considered in this work.

%
%
%
\subsection{Electronic structure}
\label{sec:struct}

The electronic-structure part of the energy is represented by an expansion in the number of
virtual photons exchanged between the electrons,
\begin{align}\label{eq:struc}
E_{\rm struc}(v) =  E_{\rm D} + E_{\rm 1phot} + E_{\rm 2phot} + E_{\rm 3phot} + E_{\rm \ge 4phot}\,,
\end{align}
where $E_{\rm D}$ is the Dirac one-electron energy; $E_{\rm 1phot}$, $E_{\rm 2phot}$, and $E_{\rm
3phot}$ are corrections due to the exchange of one, two, and three virtual photons, respectively,
and $E_{\rm \ge 4phot}$ corresponds to the exchange by four and more photons.

The Dirac ionization energy of the valence state $v$, for the point nuclear model, is given by
the well-known formula
\begin{align}\label{eq:Dirac}
E_{\rm D}(v) = \left[1+\bigg(\frac{Z\alpha}{n_v - |\kappa_v|+\sqrt{\kappa_v^2-(Z\alpha)^2}}\bigg)^2
  \right]^{-1/2} - 1\,,
\end{align}
where $n_v$ and $\kappa_v$ are the principal and the relativistic angular quantum numbers of the
state $v$, respectively. The point-nucleus Dirac energy receives a correction from the finite
nuclear size (fns), which is very small for low-$Z$ ions but becomes increasingly important as
$Z$ increases. The fns correction can be easily calculated numerically, by solving the Dirac
equation with a suitable nuclear binding potential.

The electron-structure corrections to the Dirac energy arise through the electron-electron
interaction. The relativistic operator of the electron-electron interaction depends on the energy
of the exchanged virtual photon $\omega$ and is given, in the Feynman gauge, by
\begin{equation}
  I_{\rm Feyn}(\omega) = \alpha\,
  \big( 1-\balpha_1 \cdot \balpha_2 \big)\,
  \frac{e^{i|\omega|x_{12}}}{x_{12}} \,,
\end{equation}
where $\balpha_{1}$ and $\balpha_{2}$ are vectors of Dirac matrices acting on the coordinate
$\bfx_{1}$ and $\bfx_2$, respectively, and $x_{12} = |\bm{x}_{12}| = |\bm{x}_{1}-\bm{x}_{2}|$.
The electron-electron interaction operator in the Coulomb gauge is
\begin{align}
  I_{\rm Coul}(\omega) = &\,\alpha\,
  \Bigg[ \frac1{x_{12}} - \balpha_1\cdot\balpha_2\,
   \frac{e^{i|\omega|x_{12}}}{x_{12}}
 \nonumber \\ &
   + \frac{\big(\balpha_1\cdot\bnabla_1\big) \big(\balpha_2\cdot\bnabla_2\big)}{\omega^2}
   \frac{e^{i|\omega|x_{12}}-1}{x_{12}}\Bigg]\,.
\end{align}

Despite the dependence of the electron-electron interaction operator $I$ on the choice of the
gauge, all terms of the expansion (\ref{eq:struc}) are gauge invariant, when calculated
rigorously within QED.
In the present work, we perform QED calculations of the corrections due to exchange by one and
two virtual photons, $E_{\rm 1phot}$ and $E_{\rm 2phot}$. The corrections induced by an exchange
of three or more photons are calculated within the Breit approximation, which is equivalent to
choosing the Coulomb gauge in the photon propagator and setting $\omega \to 0$.

In the following, we will extensively use  the following short-hand notations for the matrix
elements of the electron-electron interaction operator,
\begin{align}
&\ \I{a}{b}{c}{d}(\Delta)  \equiv \lbr ab| I(\Delta)|cd\rbr\,,& \\
&\ \I{a}{b}{c}{d}  \equiv \lbr ab| I_{\rm Coul}(0)|cd\rbr\,.&
\end{align}

The leading electron-structure correction comes from the exchange of one virtual photon between
the electrons. The correction due to one-photon exchange between a valence electron $v$ and a
closed shell of electron states $c$ is given by
\begin{align} \label{eq:1phot}
E_{\rm 1phot}(v) &\, =
  \sum_{\mu_c} \sum_P (-1)^P  \I{Pv}{Pc}{v}{c}(\Delta_{Pcc})
  \nonumber \\
  &\,
  \equiv \sum_{\mu_c}\big[ \I{v}{c}{v}{c}(0) - \I{c}{v}{v}{c}(\Delta_{vc})\Big]\,,
\end{align}
where $P$ is the permutation operator interchanging the one-electron states, $(PvPc) = (vc)$ or
$(cv)$, $(-1)^P$ is the sign of the permutation, $\Delta_{ab} = \vare_a -\vare_b$ is the
difference of one-electron energies, and the summation over $\mu_c$ runs over the angular
momentum projections of the core electrons. The one-photon exchange correction is relatively
simple and can be calculated to very high numerical accuracy.

The effects caused by the exchange of two photons are much more complicated than the one-photon
contribution. First rigorous QED calculations of the two-photon exchange correction started in
1990th and were performed for He-like ions
\cite{blundell:93:b,lindgren:95:pra,mohr:00:pra,andreev:01}. For Li-like ions, analogous
calculations were accomplished in
Refs.~\cite{yerokhin:00:prl,yerokhin:01:2ph,artemyev:03,kozhedub:10,sapirstein:11}. In the
present work, we extend the previous calculations described in
Refs.~\cite{yerokhin:00:prl,yerokhin:01:2ph,artemyev:03} to a greater numerical accuracy and a
larger interval of nuclear charges.

\begin{widetext}
The correction induced by the two-photon exchange between a valence electron $v$ and a closed
shell of electron states $c$ is given by \cite{yerokhin:01:2ph}
\begin{align}\label{eq:2ph}
E_{\rm 2phot}(v) = &\ \sum_{\mu_c}\sum_{P}(-1)^P
    \left.\sum_{n_1n_2}\right.^{\!\prime}
 \frac{i}{2\pi} \intinf d\omega\,
    \Bigg[
     \frac{\I{Pc}{Pv}{n_1}{n_2}(\omega)\, \I{n_1}{n_2}{c}{v}(\omega-\Delta_{Pcc})}
     {(\vare_{Pc} -\omega -u\vare_{n_1})(\vare_{Pv} +\omega -u\vare_{n_2})}
  +   \frac{\I{Pc}{n_2}{n_1}{v}(\omega)\, \I{n_1}{Pv}{c}{n_2}(\omega-\Delta_{Pcc})}
     {(\vare_{Pc} -\omega -u\vare_{n_1})(\vare_v -\omega -u\vare_{n_2})}
     \Bigg]
       \nonumber \\ &
{} +   \sum_{PQ}(-1)^{P+Q}\,
    \left.\sum_{n}\right.^{\prime}
       \frac{\I{P2}{P3}{n}{Q3}(\Delta_{P3Q3})\, \I{P1}{n}{Q1}{Q2}(\Delta_{Q1P1})}
      {\vare_{Q1}+\vare_{Q2}-\vare_{P1}-\vare_{n}} + E_{\rm red}(v)\,,
\end{align}
where $P$ and $Q$ are the permutation operators, $u \equiv 1-i0$, and the prime on the sum symbol
means that some terms are excluded from the summation (the excluded terms are ascribed to the
reducible part $E_{\rm red}$ and evaluated separately, see
Refs.~\cite{yerokhin:01:2ph,artemyev:03} for details). In Eq.~(\ref{eq:2ph}), the first part on
the right-hand side is the irreducible two-electron contribution, the second part is the
irreducible three-electron contribution (with "1", "2", and "3" numerating the three electrons,
in arbitrary order), and the third part $\Delta E_{\rm red}$ is the reducible contribution. The
explicit expression for $\Delta E_{\rm red}$ can be found in
Refs.~\cite{yerokhin:01:2ph,artemyev:03}.

The two-photon exchange correction can be greatly simplified in the MBPT approximation, which
assumes that (i) the electron-electron interaction is taken in the Breit approximation,
$I(\omega) \to I_{\rm Coul}(0)$ and (ii) the summations are performed over the positive-energy
part of the Dirac spectrum. Within this approximation, the integration over $\omega$ is performed
by the Cauchy theorem and the crossed-photon and reducible contributions vanish, yielding the
result
\begin{align}\label{eq:2ph:mbpt}
E_{\rm 2phot}^{\rm MBPT}(v) &\ = \sum_{\mu_c}\sum_{P}(-1)^P
    \left.\sum_{n_1n_2}\right.^{\!\prime(+)}
     \frac{\I{Pc}{Pv}{n_1}{n_2}\, \I{n_1}{n_2}{c}{v}}
     {\vare_c + \vare_v - \vare_{n_1} - \vare_{n_2}}
+   \sum_{PQ}(-1)^{P+Q}\,
    \left.\sum_{n}\right.^{\prime(+)}
       \frac{\I{P2}{P3}{n}{Q3}\, \I{P1}{n}{Q1}{Q2}}
      {\vare_{Q1}+\vare_{Q2}-\vare_{P1}-\vare_{n}} \,,
\end{align}
where the prime on the summation symbol means that terms with vanishing denominator are omitted
and ``$(+)$'' means that the summation is extended over the positive-energy part of the Dirac
spectrum.

The three-photon exchange correction cannot be presently calculated rigorously within QED. In the
present work we evaluate it within the MBPT approximation, where it is represented as
\cite{yerokhin:07:lilike}
\begin{align}\label{eq:3ph}
 \Delta E^{\rm MBPT}_{\rm 3ph}(v) =&\
    \sum_{\mu_c}\, \sum_P (-1)^P\,
    \left.\sum_{n_1\ldots n_4}\right.^{\!\!\!(+)}
         \Xi_1 \,\frac{\I{Pv}{Pc}{n_1}{n_2}\, \I{n_1}{n_2}{n_3}{n_4}\,
         \I{n_3}{n_4}{v}{c}}
         {(\vare_c+\vare_v-\vare_{n_1}-\vare_{n_2})(\vare_c+\vare_v-\vare_{n_3}-\vare_{n_4})}
   \nonumber \\ &
{} +  \sum_{PQ} (-1)^{P+Q}\,   \sum_{n_1 n_2 n_3}{\!\!\!}^{^{(+)}}\, \Xi_1 \,\left[
         \frac{2\,\I{P2}{P3}{n_1}{Q3}\,
    \I{P1}{n_1}{n_2}{n_3}\, \I{n_2}{n_3}{Q1}{Q2}}
          {(\vare_{Q1}+\vare_{Q2}-\vare_{P1}-\vare_{n_1})(\vare_{Q1}+\vare_{Q2}-\vare_{n_2}-\vare_{n_3})}
 \right. \nonumber \\ &
{}    + \frac{\I{P1}{P2}{n_1}{n_2}\,
    \I{n_2}{P3}{n_3}{Q3}\, \I{n_1}{n_3}{Q1}{Q2}}
          {(\vare_{P1}+\vare_{P2}-\vare_{n_1}-\vare_{n_2})(\vare_{Q1}+\vare_{Q2}-\vare_{n_1}-\vare_{n_3})}
       \left.
{}+         \frac{\I{P2}{P3}{n_1}{n_2}\,
    \I{P1}{n_1}{n_3}{Q2}\, \I{n_3}{n_2}{Q1}{Q3}}
          {(\vare_{P2}+\vare_{P3}-\vare_{n_1}-\vare_{n_2})(\vare_{Q1}+\vare_{Q3}-\vare_{n_2}-\vare_{n_3})}
\right]\,,
 \nonumber \\
\end{align}
\end{widetext}
where the operator $\Xi_1$ acts on energy denominators $\Delta_1$, $\Delta_2$ as following
\begin{eqnarray} \label{eqII7}
\Xi_1 \, \frac{X}{\Delta_1\,\Delta_2} = \left\{
   \begin{array}{cl}
 \displaystyle       \frac{ X}{ \Delta_1\,\Delta_2}\,,
        & \mbox{if}\         \Delta_1\ne0\,,\Delta_2\ne0\,,\\[0.25cm]
        \displaystyle -\frac{ X}{2\,\Delta_1^2}\,, &\mbox{if}\
                             \Delta_1\ne0\,,\Delta_2=0\,,\\[0.25cm]
        \displaystyle  -\frac{X}{2\,\Delta_2^2}\,, &\mbox{if}\
                             \Delta_1=0\,,\Delta_2\ne0\,,\\[0.25cm]
        0\,, &\mbox{if}\
                             \Delta_1=0\,,\Delta_2=0\,.\\
   \end{array}
   \right.
\end{eqnarray}

The correction induced by the exchange of four and more photons, $E_{\ge \rm 4phot}$, is too
complicated to be calculated by perturbation theory. In the present work we extract this
correction from the NRQED results, which account for all orders in $1/Z$ but only the leading
order in $\Za$; the corresponding calculation is described in Sec.~\ref{sec:NRQED}.

\subsection{Radiative QED}
\label{sec:radqed}

The radiative QED contribution to the fs splitting is represented as an expansion in the number
of virtual photons exchanged between the electrons (with the expansion parameter $1/Z$),
\begin{align}\label{eq:radqed}
E_{\rm rad} =  E_{\rm QEDhydr} + E_{\rm QEDscr1} + E_{\rm QEDscr2} + E_{\rm QEDscr3+}\,,
\end{align}
where $E_{\rm QEDhydr}$ is the hydrogenic QED correction, $E_{\rm QEDscr1}$ is the screening QED
correction with one electron-electron interaction, $E_{\rm QEDscr2}$ is the screening QED
correction with two electron-electron interactions, and $E_{\rm QEDscr3+}$ contains three and
more electron-electron interactions.

The one-electron QED contribution $E_{\rm QEDhydr}$ is presently well established, see, e.g., a
recent review \cite{yerokhin:15:Hlike}; it will be taken from the literature in this work. The
first-order $1/Z^1$ screening QED correction $E_{\rm QEDscr1}$ was calculated for Li-like ions in
Refs.~\cite{yerokhin:99:sescr,artemyev:99,yerokhin:05:OS,sapirstein:01:lamb,kozhedub:10,sapirstein:11};
numerical results for this correction will also be taken from the literature.

\begin{table}
\caption{
Comparison of different approximate methods with the rigorous QED calculations
\cite{yerokhin:99:sescr,artemyev:99,yerokhin:05:OS}
of the first-order $1/Z^1$ QED screening correction, in units $\alpha^2(\Za)^3$.
\label{tab:ammqed}}
\begin{ruledtabular}
\begin{tabular}{lw{4.6}w{4.6}w{4.6}w{4.8}}
 $Z$
    & \multicolumn{1}{c}{amm}
      & \multicolumn{1}{c}{MQED}
          & \multicolumn{1}{c}{amm+MQED}
                  & \multicolumn{1}{c}{Full QED}
 \\\hline\\[-5pt]
%
 12   &  -0.0658  &   -0.0362  &   -0.0618    &    -0.0616\,(14) \\
 16   &  -0.0664  &   -0.0341  &   -0.0601    &    -0.0590\,(9) \\
 18   &  -0.0667  &   -0.0330  &   -0.0592    &    -0.0579\,(4) \\
 20   &  -0.0671  &   -0.0319  &   -0.0582    &    -0.0566\,(3) \\
 30   &  -0.0699  &   -0.0253  &   -0.0529    &    -0.0501\,(3) \\
 32   &  -0.0706  &   -0.0238  &   -0.0517    &    -0.0486\,(4) \\
 40   &  -0.0741  &   -0.0173  &   -0.0465    &    -0.0422\,(2) \\
 50   &  -0.0803  &   -0.0075  &   -0.0387    &    -0.0325\,(2) \\
 54   &  -0.0835  &   -0.0029  &   -0.0350    &    -0.0281\,(2) \\
 60   &  -0.0893  &    0.0049  &   -0.0285    &    -0.0202\,(2) \\
 66   &  -0.0967  &    0.0142  &   -0.0209    &    -0.0113\,(2) \\
 70   &  -0.1028  &    0.0213  &   -0.0149    &    -0.0041\,(1) \\
 74   &  -0.1101  &    0.0294  &   -0.0080    &     0.0037\,(2) \\
 80   &  -0.1241  &    0.0439  &    0.0043    &     0.0182\,(1) \\
 83   &  -0.1329  &    0.0523  &    0.0115    &     0.0266\,(1) \\
 90   &  -0.1601  &    0.0761  &    0.0319    &     0.0501\,(1) \\
 92   &  -0.1702  &    0.0842  &    0.0389    &     0.0581\,(1) \\
100   &  -0.2277  &    0.1240  &    0.0728    &     0.0974\,(1) \\
\end{tabular}
\end{ruledtabular}
\end{table}
We now concentrate on the second-order $1/Z^2$ screening QED contribution $E_{\rm QEDscr2}$. At
present, it is not possible to calculate this correction rigorously to all orders in $\Za$. In
this work, we will calculate it by an approximate relativistic method which is exact to the
leading order in $\Za$ and accounts for the dominant part of the higher-order $\Za$ terms.

It is well-known \cite{hegstrom:73} that, to the leading order in $\Za$, the radiative QED
effects in the fs splitting are described by the electron anomalous magnetic moment (amm). In the
absence of external fields, the electron amm induces the following two additions to the Dirac
Hamiltonian of a few-electron atom \cite{hegstrom:73,glazov:04:pra},
\begin{align} \label{eq:amm1}
H_{\rm amm,1} = &\ \kappa\, \frac{\Za}{4}\,(-i)\,\sum_a \beta_a\,\frac{\balpha_a\cdot\bfr_a}{r_a^3}\,,
\\
H_{\rm amm,2} = &\ \kappa\, \frac{\alpha}{4}\,\sum_{a < b}
  \beta_a\,\Big(
     i\,\frac{\balpha_a\cdot\bfr_{ab}}{r_{ab}^3}
  -\bm{\Sigma}_a\cdot \frac{\balpha_b \times \bfr_{ab}}{r_{ab}^3}
     \Big) \,,
      \label{eq:amm2}
\end{align}
where $a$ and $b$ numerate the electrons in the atom, $\kappa = g_e-2 = \nicefrac{\alpha}{\pi} +
O(\alpha^2)$, $g_e$ is the $g$-factor of the free electron, $\beta_a$ and $\balpha_a$ are the
Dirac matrices acting on $a$th electron, and
\begin{align}
\bm{\Sigma} = \Dmatrix{\bsigma}{0}{0}{\bsigma} \,,
\end{align}
with $\bsigma$ being a vector of Pauli matrices.

The effective amm Hamiltonian $H_{\rm amm} = H_{\rm amm, 1} + H_{\rm amm, 2}$ yields a good
description of the radiative QED effects for low-$Z$ ions, but the accuracy deteriorates quickly
when $Z$ increases. We will correct this with help of the model QED (MQED) operator $h_{\rm
MQED}$ introduced in Ref.~\cite{shabaev:13:qedmod}. In order to avoid double counting, we
subtract from $h_{\rm MQED}$ the part already accounted for by the amm Hamiltonian. Specifically,
we make the replacement
\begin{align}
&\frac12 \lbr \psi_j | \Sigma(\vare_j)+\Sigma(\vare_l)|\psi_l\rbr
 \to
  \nonumber \\
& \ \ \  \ \ \ \ \ \ \ \
\frac12 \lbr \psi_j | \Sigma(\vare_j)+\Sigma(\vare_l)|\psi_l\rbr
 -
 \lbr \psi_j | H_{\rm amm,1}|\psi_l\rbr
\end{align}
in the definition of the MQED operator (where $\Sigma(\vare)$ is the self-energy operator), see
Eq.~(17) of Ref.~\cite{shabaev:13:qedmod}. We will denote the amm-subtracted MQED operator by
$h_{\rm MQED}'$.

In this work we will calculate the second-order QED screening correction $E_{\rm QEDscr2}$ by
using the standard Rayleigh-Schr\"odinger perturbation theory to the second order in the
electron-electron interaction and to first order in the effective Hamiltonian $H_{\rm amm+MQED}$,
\begin{align}\label{eq:Hammqed}
H_{\rm amm+MQED} = H_{\rm amm, 1} + H_{\rm amm, 2} +h_{\rm MQED}' \equiv U + W\,.
\end{align}
The operators $U$ and $W$ introduced in the right-hand-side of the above equation incorporate the
one-electron part ($H_{\rm amm, 1} +h_{\rm MQED}'$) and the two-electron part ($H_{\rm amm, 2} $)
of the effective Hamiltonian, respectively.

Before calculating the second-order screening QED effect, we need to check the accuracy of the
approximate method we devised. We do this by applying this approximation to the first-order
screening QED correction and comparing the obtained results with those delivered by the rigorous
QED calculations.

The $1/Z^1$  correction induced by the one-electron operator $U$ is obtained as a first-order (in
$U$) perturbation of the one-photon exchange correction (\ref{eq:1phot}), which is (after
dropping the frequency-dependent terms)
\begin{align}
    E_{{\rm 1phot},U} = 2 \sum_{\mu_c}\sum_P (-1)^P \left(
    \I{Pv}{Pc}{\delta v}{c}+\I{Pv}{Pc}{v}{\delta c} \right) \,,
\end{align}
where
\begin{align}
|\delta a\rbr = \left.\sum_{n}\right.^{\prime} \frac{|n\rbr\,U_{na}}{\vare_a-\vare_n}\,,
\end{align}
and $U_{ab} \equiv \lbr a|U|b\rbr$. The $1/Z^1$  correction induced by the two-electron operator
$W$ is just
\begin{align}
    E_{{\rm 1phot},W} =  \sum_{\mu_c}\sum_P (-1)^P
    \W{Pv}{Pc}{v}{c} \,,
\end{align}
where $\W{a}{b}{c}{d} \equiv \lbr ab|W|cd\rbr$.

Table~\ref{tab:ammqed} presents results of our test calculations of the first-order $1/Z^1$ QED
screening correction performed by three approximate methods and compares them with results
obtained by the full QED treatment. The column ``amm'' lists results obtained with the amm
operator $H_{\rm amm}$, the column ``MQED'' displays results obtained with the standard MQED
operator $h_{\rm MQED}$, whereas the column ``amm+MQED'' shows results obtained with the combined
operator (\ref{eq:Hammqed}).

We observe that the approach based on the amm Hamiltonian works well only in the low-$Z$ region
but fails for high values of $Z$, not reproducing even the overall sign of the effect. The
standard MQED operator yields the order of magnitude and the sign of the exact QED screening
correction, but the quantitative agreement is not very good. In contrast, the combined
``amm+MQED'' approach demonstrates a significantly improved agreement with the rigorous QED
treatment as compared to the both other methods.

\begin{widetext}

We now turn to the second-order $1/Z^2$ screening QED effect. The $1/Z^2$  correction induced by
the one-electron operator $U$ can be derived as a first-order (in $U$) perturbation of the
two-photon exchange correction in the MBPT approximation, given by Eq.~(\ref{eq:2ph:mbpt}). It
consists of 3 parts that are induced by perturbations of the wave functions (``wf"), binding
energies (``en"), and propagators (``ver"), respectively,
\begin{equation} \label{eqII13}
E_{\rm 2phot,U} = E_{\rm 2phot,wf}+ E_{\rm 2phot,en}+
E_{\rm 2phot,ver}\,.
\end{equation}
The corresponding parts are given by
\begin{eqnarray} \label{eqII14}
     E_{\rm 2phot,wf} &=& 2\,\sum_{\mu_c}\sum_P
    (-1)^P\,
    \left.\sum_{n_1n_2}\right.^{\!\prime(+)}
        \frac{\I{Pv}{Pc}{n_1}{n_2}\, \big( \I{n_1}{n_2}{\delta v}{c}+ \I{n_1}{n_2}{v}{\delta c}\big)}
                      {\vare_c+\vare_v-\vare_{n_1}-\vare_{n_2}}
    \nonumber \\ &&
{} + 2\,  \sum_{PQ}(-1)^{P+Q}\,
    \left.\sum_{n}\right.^{\prime(+)}
       \frac{\I{P2}{P3}{n}{Q3}\, \big( \I{\delta P1}{n}{Q1}{Q2}+
      \I{P1}{n}{\delta Q1}{Q2}+ \I{P1}{n}{Q1}{\delta Q2}\big)}
      {\vare_{Q1}+\vare_{Q2}-\vare_{P1}-\vare_{n}}\,,
\end{eqnarray}
\begin{eqnarray} \label{eqII15}
     E_{\rm 2phot,en} &=&
 -(\U{v}{v}+\U{c}{c})\, \sum_{\mu_c}\sum_P
    (-1)^P\,
    \left.\sum_{n_1n_2}\right.^{\!\prime(+)}
        \frac{\I{Pv}{Pc}{n_1}{n_2}\,\I{n_1}{n_2}{v}{c}}{(\vare_c+\vare_v-\vare_{n_1}-\vare_{n_2})^2}
    \nonumber \\ &&
  {}  - \sum_{PQ}(-1)^{P+Q}\, (\U{Q1}{Q1}+\U{Q2}{Q2}-\U{P1}{P1})\,
    \left.\sum_{n}\right.^{\prime(+)}
      \frac{\I{P2}{P3}{n}{Q3}\,\I{P1}{n}{Q1}{Q2}}{(\vare_{Q1}+\vare_{Q2}-\vare_{P1}-\vare_{n})^2}\,,
\end{eqnarray}
\begin{eqnarray} \label{eqII16}
     E_{\rm 2phot,ver} &=&
 \sum_{\mu_c}\sum_P
    (-1)^P\,
    \left.\sum_{n_1n_2n_3}\right.^{\!\!\!\!(+)}
    \Xi_2\,
        \frac{\I{Pv}{Pc}{n_1}{n_2}}{\vare_c+\vare_v-\vare_{n_1}-\vare_{n_2}}
  \left(
        \frac{\U{n_1}{n_3}\,\I{n_3}{n_2}{v}{c}}{
              \vare_c+\vare_v-\vare_{n_3}-\vare_{n_2}}
 +\frac{\U{n_2}{n_3}\,\I{n_1}{n_3}{v}{c}}{
              \vare_c+\vare_v-\vare_{n_1}-\vare_{n_3}}
    \right)
    \nonumber \\ &&
    + \sum_{PQ}(-1)^{P+Q}\,
    \left.\sum_{n_1n_2}\right.^{\!(+)}
    \Xi_2\,  \frac{\I{P2}{P3}{n_1}{Q3}\,\U{n_1}{n_2}\,\I{P1}{n_2}{Q1}{Q2}}
    {(\vare_{Q1}+\vare_{Q2}-\vare_{P1}-\vare_{n_1})(\vare_{Q1}+\vare_{Q2}-\vare_{P1}-\vare_{n_2})}\,,
\end{eqnarray}
 where the operator $\Xi_2$ acts on energy denominators $\Delta_1$,
$\Delta_2$ as following:
\begin{eqnarray} \label{eqII17}
\Xi_2 \, \frac{X}{\Delta_1\,\Delta_2} = \left\{
   \begin{array}{cl}
 \displaystyle       \frac{ X}{ \Delta_1\,\Delta_2}\,,
        & \mbox{if}\         \Delta_1\ne0\,,\Delta_2\ne0\,,\\[0.3cm]
        \displaystyle -\frac{ X}{\Delta_1^2}\,, &\mbox{if}\
                             \Delta_1\ne0\,,\Delta_2=0\,,\\[0.3cm]
        \displaystyle  -\frac{X}{\Delta_2^2}\,, &\mbox{if}\
                             \Delta_1=0\,,\Delta_2\ne0\,,\\[0.3cm]
        0\,, &\mbox{if}\
                             \Delta_1=0\,,\Delta_2=0\,.\\
   \end{array}
   \right.
\end{eqnarray}
We note that similar formulas appeared in a slightly different context in
Ref.~\cite{yerokhin:07:lilike} ({\em cf.} Eqs.~(32)-(35) of that work).

The $1/Z^2$  correction induced by the two-electron operator $W$ is given by
\begin{eqnarray}
     E_{\rm 2phot,W} &=& \sum_{\mu_c}\sum_P
    (-1)^P\,
    \left.\sum_{n_1n_2}\right.^{\!\prime(+)}
        \frac{\I{Pv}{Pc}{n_1}{n_2}\, \W{n_1}{n_2}{v}{c} + \W{Pv}{Pc}{n_1}{n_2}\, \I{n_1}{n_2}{v}{c}}
                      {\vare_c+\vare_v-\vare_{n_1}-\vare_{n_2}}
    \nonumber \\ &&
{} +   \sum_{PQ}(-1)^{P+Q}\,
    \left.\sum_{n}\right.^{\prime(+)}
       \frac{\I{P2}{P3}{n}{Q3}\,\W{P1}{n}{Q1}{Q2}+ \W{P2}{P3}{n}{Q3}\,\I{P1}{n}{Q1}{Q2}}
      {\vare_{Q1}+\vare_{Q2}-\vare_{P1}-\vare_{n}}\,.
\end{eqnarray}

\end{widetext}

\subsection{Nuclear recoil}

The nuclear recoil contribution is represented in this work as a sum of two parts,
\begin{align}
E_{\rm rec} = E_{\rm rec}^{\,\rm oneel} +  E_{\rm rec}^{\,\rm fewel}\,,
\end{align}
where the first part is the one-electron (hydrogenic) contribution and the second part is the
few-body contribution. The one-electron contribution is presently well established, see, e.g., a
recent review \cite{yerokhin:15:Hlike}, and is taken from the literature. The few-body recoil
contribution will be evaluated to the leading order in $\Za$ within the NRQED approach in next
Section.

\begin{table*}[t]
\caption{Numerical results for the $\alpha^4$, $\alpha^5$, and $\alpha^4(m/M)$ corrections to the
fine structure of the $2P$ state of Li-like ions. ${\cal E}^{(i,j)}$
are defined by Eq.~(\ref{eq:nrqed}).
\label{tab:NRQED}}
\begin{ruledtabular}
\begin{tabular}{lw{4.16}w{4.16}w{4.16}}
 $Z$
  & \multicolumn{1}{c}{${\cal E}^{(4,0)}/Z^4$ }
  & \multicolumn{1}{c}{${\cal E}^{(5,0)}/Z^4$ }
  & \multicolumn{1}{c}{${\cal E}^{(4,1)}/Z^4$ }
 \\\hline\\[-5pt]
3  & 0.000\,353\,014\,9\,(1) & 0.000\,169\,064\,948\,(1) & -0.001\,074\,97\,(2) \\
4  & 0.002\,190\,425\,2\,(5) & 0.000\,865\,489\,89\,(10) & -0.003\,891\,58\,(4) \\
5  & 0.004\,653\,760\,1\,(3) & 0.001\,713\,290\,95\,(7)  & -0.007\,034\,07\,(2) \\
6  & 0.007\,054\,738\,0\,(2) & 0.002\,505\,289\,60\,(3)  & -0.010\,032\,02\,(2) \\
7  & 0.009\,196\,952\,5\,(7) & 0.003\,195\,699\,19\,(15) & -0.012\,717\,88\,(4) \\
8  & 0.011\,057\,748\,2\,(1) & 0.003\,786\,733\,68\,(2)  & -0.015\,068\,04\,(1) \\
9  & 0.012\,664\,475\,3\,(1) & 0.004\,292\,004\,93\,(4)  & -0.017\,111\,05\,(1) \\
10 & 0.014\,054\,569\,0\,(1) & 0.004\,725\,993\,9\,(2)   & -0.018\,888\,65\,(2) \\
11 & 0.015\,263\,384\,0\,(1) & 0.005\,101\,316\,3\,(3)   & -0.020\,441\,55\,(1) \\
12 & 0.016\,321\,112\,5\,(1) & 0.005\,428\,312\,6\,(2)   & -0.021\,805\,51\,(1) \\
13 & 0.017\,252\,626\,(3)    & 0.005\,715\,289\,(8)      & -0.023\,010\,5\,(5)  \\
\end{tabular}
\end{ruledtabular}
\end{table*}

\section{Non-relativistic QED}
\label{sec:NRQED}

In the nonrelativistic quantum electrodynamics (NRQED) framework, the fs splitting of light atoms
is represented by an expansion in powers of the fine-structure constant $\alpha$ and the
electron-to-nucleus mass ratio $m/M$ \cite{puchalski:09,puchalski:14},
\begin{align} \label{eq:nrqed}
E_{\rm NRQED}
  =   &\, \alpha^4 \left[ {\cal E}^{(4,0)} + \frac{m}{M}\,{\cal E}^{(4,1)}
     +  \alpha\, {\cal E}^{(5,0)}  + \ldots \right]\,.
\end{align}
Here, the first superscript of the expansion terms ${\cal E}^{(i,j)}$ indicates the order in
$\alpha$, whereas the second superscript shows the order in $m/M$. Each term of the NRQED
expansion is represented as an expectation value of some effective Hamiltonian on the
nonrelativistic atomic wave function and thus accounts for the nonrelativistic electron-electron
interaction (i.e., the parameter $1/Z$) to all orders.

The leading term of the NRQED expansion of the fs interval is given by the difference of the
expectation values of the spin-dependent Breit Hamiltonian, ${\cal E}^{(4,0)} = \lbr
H^{(4,0)}\rbr_{J = \nicefrac32}- \lbr H^{(4,0)}\rbr_{J = \nicefrac12}$. The spin-dependent part
of the Breit Hamiltonian is (in atomic units)
\begin{align}
H^{(4,0)} = & \sum_a \frac{Z}{2\,r_a^3}\,\vec{s}_a\cdot \vec{r}_a\times \vec{p}_a
 \nonumber \\ &
 + \sum_{a \neq b} \frac{1}{2\,r_{ab}^3}\,\vec{s}_a\cdot \vec{r}_{ab}\times \big(2\vec{p}_b
  - \vec{p}_a\big) \,,
\end{align}
where $a$ and $b$ numerate electrons in the atom, $\vec{r}_{ab} = \vec{r}_a-\vec{r}_b$,
$\vec{p}_a$ is the electron momentum, and $\vec{s}_a$ is the electron-spin operator.

The spin-dependent $m\alpha^4$ recoil correction for a state with the total angular momentum $J$
is given by (in atomic units)
\begin{align}
{\cal E}^{(4,1)}_{J} = & \Big< H^{(4,0)}\, \frac1{(E-H)'}\, H^{(2,1)}\Big>_{\!J}
 \nonumber \\
 & + \Big< \sum_{ab} \frac{Z}{r_a^3}\,\vec{s}_a\cdot \vec{r}_a\times \vec{p}_b\Big>_{\!J}\,,
\end{align}
where $H^{(2,1)}$ is the recoil operator of order $m\alpha^2$,
\begin{align}
H^{(2,1)} = \frac12\, \vec{P}^2 = \frac12\, \Big(-\sum_{a}\vec{p}_a\Big)^2\,,
\end{align}
and $\vec{P}$ is the nuclear momentum.

The leading QED contribution to the fs interval is induced by the Hamiltonian $H^{(5,0)}$, ${\cal
E}^{(5,0)} = \lbr H^{(5,0)}\rbr_{J=\nicefrac32}-\lbr H^{(5,0)}\rbr_{J=\nicefrac12}$, where (in
atomic units)
\begin{align}
H^{(5,0)} = & \sum_a \frac{Z}{2\pi\,r_a^3}\,\vec{s}_a\cdot \vec{r}_a\times \vec{p}_a
 \nonumber \\ &
 + \sum_{a \neq b} \frac{1}{2\pi\,r_{ab}^3}\,\vec{s}_a\cdot \vec{r}_{ab}\times\big(\vec{p}_b
  - \vec{p}_a\big) \,.
\end{align}

In the present work we calculate the corrections ${\cal E}^{(4,0)}$, ${\cal E}^{(5,0)}$, and
${\cal E}^{(4,1)}$ for the series of nuclear charges $Z = 3$--$13$. The computational scheme and
numerical details are described in Ref.~\cite{puchalski:09,puchalski:14}. Our numerical results
are presented in Table~\ref{tab:NRQED}.

In order to combine the NRQED results with those obtained within the $1/Z$-exansion method in
Sec.~\ref{sec:allorder}, we represent the numerical results listed in Table~\ref{tab:NRQED} in
the form of the $1/Z$ expansion,
\begin{align}\label{eq:Zexp1}
{\cal E}^{(4,0)} = &\ Z^4 \sum_{i = 0}^{\infty} \frac{C_{i,4}}{Z^i}\,,
 \\
{\cal E}^{(5,0)} = &\ Z^4 \sum_{i = 0}^{\infty} \frac{D_{i,5}}{Z^i}\,,
 \\
{\cal E}^{(4,1)} = &\ Z^4 \sum_{i = 0}^{\infty} \frac{R_{i,4}}{Z^i}\,.
\label{eq:Zexp2}
\end{align}
Here and in what follows, we adopt the following notations for the expansion coefficients
$C_{i,j}$, $D_{i,j}$, $R_{i,j}$: the first index $i$ corresponds to the order in $1/Z$, whereas
the second index $j$ indicates the order in $\alpha$.

The first coefficients of the expansions are known analytically,
\begin{align}
&C_{0,4} =  \frac1{32}\,, \ \ \
D_{0,5} =  \frac1{32\pi}\,,&\\
&R_{0,4} =  -\frac{1}{32} + \frac{2^8}{3^9}\big( 3\ln \frac32 - 2\big)\,,&
\end{align}
where $C_{0,4}$ comes from the $\Za$ expansion of the Dirac energy (\ref{eq:Dirac}), $D_{0,5}$
comes from the one-loop self-energy (see, e.g., Eq.~(38) of Ref.~\cite{mohr:16:codata}), whereas
the $R_{0,4}$ coefficient was derived in Ref.~\cite{shabaev:94:rec}. The coefficients $C_{1,4}$
and $C_{2,4}$ will be numerically evaluated in the next Section, by calculating the one-photon
and two-photon exchange corrections and fitting their $Z \to 0$ and $\alpha \to 0$ limit. The
other coefficients in Eqs.~(\ref{eq:Zexp1})-(\ref{eq:Zexp2}) are approximately obtained by
fitting the numerical results from Table~\ref{tab:NRQED}.

\section{Calculational details and results}

\subsection{Electronic structure}

Table~\ref{tab:struct} presents results of our numerical calculations of individual
electron-structure contributions. The column labeled ``Dirac'' shows the Dirac one-electron
energies $E_D$. The uncertainties of $E_D$, appearing for high-$Z$ ions, are due to the finite
nuclear size effect. The $\Za$ expansion of the Dirac fs splitting follows from
Eq.~(\ref{eq:Dirac}),
\begin{align}
E_{D} = (\Za)^4 \left[ C_{0,4} + (\Za)^2\, C_{0,6} + (\Za)^4\, C_{0,8} + \ldots\right]\,,
\end{align}
where $C_{0,4} = \nicefrac{1}{32}$, $C_{0,6} = \nicefrac{5}{256}$, etc.

The next column labeled ``1-ph'' contains results for the one-photon exchange correction. Its
calculation is relatively straightforward and can be preformed up to arbitrary numerical
accuracy. The $\Za$ expansion of the one-photon exchange correction for the fs splitting is of
the form
\begin{align}
E_{\rm 1phot} = \alpha (\Za)^3 \left[ C_{1,4} + (\Za)^2\,C_{1,6} + (\Za)^4\,C_{1,8} + \ldots \right]\,.
\end{align}
While our numerical calculation accounts for all orders in $\Za$, we also determine values of the
first two expansion coefficients by fitting our all-order results, obtaining $C_{1,4} =
-0.218\,109\,12$ and $C_{1,6} = -0.194\,777$.

The two-photon exchange correction is calculated in the present work rigorously within QED, by
the method described in the previous investigations \cite{yerokhin:01:2ph,artemyev:03}. The Dirac
spectrum is represented by using the dual kinetic balance (DKB) method \cite{shabaev:04:DKB} with
$N = 85$ $B$-spline basis functions. The partial-wave expansion was extended up to $|\kappa_{\rm
max}| = 20$, with the remaining tail of the expansion estimated by a least-square fitting in
$1/|\kappa|$. The direct numerical calculations were performed for $Z > 13$.

In order to obtain results for the two-photon exchange correction in the low-$Z$ region, we fit
our numerical values to the form of the $\Za$ expansion,
\begin{align} \label{eq:2phZa}
E_{\rm 2phot} = &\, \alpha^2\, (\Za)^2 \Big[ C_{2,4} + (\Za)^2\,C_{2,6} +
 \ldots \Big]\,.
\end{align}
The leading expansion coefficient $C_{2,4}$ is evaluated separately, by two different methods.
First, we obtain it by fitting the $1/Z$ expansion of the $m\alpha^4$ NRQED results obtained in
Sec.~\ref{sec:NRQED}. Second, we get it by fitting the $Z\to 0$ and $\alpha \to 0$ limit of the
two-photon exchange correction in the MBPT approximation (\ref{eq:2ph:mbpt}). Both methods yield
consisting results, but the second is more accurate. We therefore fix the coefficient as $C_{2,4}
= 0.497\,88$. With the leading coefficient $C_{2,4}$ fixed in this way, the higher-order
coefficients are obtained by fitting our numerical all-order results. In particular, we obtain
the next-order coefficient as $C_{2,6} = 0.75$.

Our numerical results for the two-photon exchange correction are presented in
Table~\ref{tab:struct}. For convenience, we separate them into two parts. The first, dominant
part is delivered by the MBPT approximation, see Eq.~(\ref{eq:2ph:mbpt}). The second, much
smaller part is the deviation of the full QED result from the MBPT value. For $Z > 13$, the
listed QED values are obtained by a direct calculation. For $Z \le 13$, the listed values are
obtained by fitting.

The three-photon exchange correction is evaluated within the MBPT approximation, according to
Eq.~(\ref{eq:3ph}). The scheme of the calculation mainly follows that of
Ref.~\cite{yerokhin:07:lilike}. However, Ref.~\cite{yerokhin:07:lilike} included the Breit
interaction up to first order only, whereas here we include in addition the exchange by two and
three Breit photons. The reason is that the inclusion of the second-order Breit exchange
significantly improves the agreement between MBPT and QED for the two-photon exchange correction
to the fs splitting.

The summations over the Dirac spectrum in the three-photon exchange correction was performed by
using the DKB method \cite{shabaev:04:DKB} with $B$-spline basis functions. The number of
$B$-splines in the basis was $N = 50$ for the three-electron part and $N = 40$ for the
two-electron part. The extrapolation of the double partial wave expansion was performed as
described in Ref.~\cite{yerokhin:07:lilike}, with the number of partial waves $l_1 = 8$ for the
first summation and $l_2 = 12$ for the second summation.

Direct numerical calculations of the three-photon exchange correction were performed for $Z \ge
20$. For lower values of $Z$, the accuracy of the numerical evaluation gradually deteriorates, so
we obtain results for this correction by fitting. Specifically, we fit our numerical results to
the form of the $\Za$-expansion
\begin{align}\label{eq:3phZa}
E_{\rm 3phot} = &\, \alpha^3\, (\Za) \left[ C_{3,4} + (\Za)^2\,C_{3,6} +
 \ldots \right]\,,
\end{align}
with the leading coefficient $C_{3,4} = -0.3681$ obtained by fitting the $1/Z$ expansion of the
NRQED results in Sec.~\ref{sec:NRQED}.  We obtain the next-order coefficient (in the MBPT
approximation) as $C_{3,6} = -1.4$.

Numerical results for the three-photon exchange correction are presented in
Table~\ref{tab:struct}, in the column labeled ``3-ph''. The uncertainty of this correction comes
mainly from unknown QED effects beyond the MBPT approximation. We estimate it by taking the
relative value of the QED-MBPT difference for the two-photon exchange correction and multiplying
it by the extension factor of 4.

The correction induced by the exchange of four and more photons $E_{\rm \ge4phot}$ is obtained
from the NRQED calculations described in Sec.~\ref{sec:NRQED}.  Direct NRQED calculations were
performed for $Z \le 13$. For these nuclear charges, we obtain $E_{\rm \ge4phot}$ by subtracting
the first terms of the $1/Z$ expansion from the $m\alpha^4$ NRQED contribution listed in
Table~\ref{tab:NRQED},
\begin{align}\label{eq:4phZa}
E_{\rm \ge4phot} = \alpha^4 {\cal E}^{(4,0)} - (\Za)^4\left[ C_{0,4} + \frac{C_{1,4}}{Z}
 + \frac{C_{2,4}}{Z^2}  + \frac{C_{3,4}}{Z^3}\right]\,.
\end{align}
We note that numerical uncertainties of the coefficients $C_{2,4}$ and $C_{3,4}$ do not
contribute to the uncertainty of the total electron-structure contribution for $Z \le 13$, since
the same coefficients used in Eqs.~(\ref{eq:2phZa}), (\ref{eq:3phZa}) and (\ref{eq:4phZa}) cancel
each other when the sum of these equations is evaluated. For $Z > 13$, we obtain $E_{\rm
\ge4phot}$ by fitting the $1/Z$ expansion of numerical results for ${\cal E}^{(4,0)}$ listed in
Table~\ref{tab:NRQED}.

Our results for $E_{\rm \ge4phot}$ are presented in Table~\ref{tab:struct}, in the column labeled
``$\ge$4-ph''. The indicated numerical uncertainty takes into account uncalculated QED effects of
order $m\alpha^6$ and higher and the uncertainty of the fit for $Z > 13$. The uncalculated
effects are estimated by taking the relative value of the deviation of the full QED results for
the two-photon exchange correction from the $m\alpha^4$ contribution induced by the coefficient
$C_{2,4}$, and multiplying it by a conservative factor of 2.

Table~\ref{tab:struct} summarizes our total numerical values of the electron-structure
contribution to the $2p_{3/2}$-$2p_{1/2}$ fs splitting in Li-like ions and compares them with
results obtained by other methods. We observe that for $Z \leq 6$, our results essentially
coincide with the $m\alpha^4$ NRQED values. The reason is that the $1/Z$ expansion, used in the
present work for calculating the higher-order QED effects, breaks down for low $Z$, with
individual $1/Z$-expansion terms cancelling each other to a great extent. For larger values of
$Z$, the convergence of the $1/Z$ expansion improves; the higher-order QED effects also become
increasingly more important, moving our results further away from the NRQED values.

For $Z\ge 10$, we compare our results with the previous {\em ab initio} QED calculation by
Kozhedub {\em et al.}~\cite{kozhedub:10}. The agreement between the calculations is excellent,
but our results are more accurate, most notably in the low-$Z$ region, due to a more complete
inclusion of many-photon exchange effects.

\subsection{Radiative QED}

We now turn to the radiative QED part, which is represented by a sum of several terms, as given
by Eq.~(\ref{eq:radqed}). The first term on the right-hand-side of Eq.~(\ref{eq:radqed}), $E_{\rm
QEDhydr}$, is due to one-electron QED effects. They were recently reviewed in
Ref.~\cite{yerokhin:15:Hlike}, so we obtain $E_{\rm QEDhydr}$ from data tabulated in that work,
adding together the one-loop and two-loop QED effects. The $\Za$ expansion of this contribution
is
\begin{align}
E_{\rm QEDhydr} = &\ \alpha (\Za)^4\,\Big[D_{0,5}
  \nonumber \\ &
+ (\Za)^2\ln(\Za)\,D^{\rm log}_{0,7} + (\Za)^2 \,D_{0,7} + \ldots\Big]\,,
\end{align}
where $D_{0,5} = \nicefrac1{32\pi}$, $D_{0,7}^{\rm log} = \nicefrac1{16\pi}$
\cite{mohr:16:codata}. The other terms on the right-hand-side of Eq.~(\ref{eq:radqed}) are due to
the electron-electron interaction; they are referred to as the screening QED corrections.

The first-order $1/Z$ screening QED correction $E_{\rm QEDscr1}$ was calculated for Li-like ions
in a series of investigations
\cite{yerokhin:99:sescr,artemyev:99,yerokhin:05:OS,sapirstein:01:lamb,kozhedub:10,sapirstein:11}.
The data reported in these studies are not fully sufficient for our present needs, because of a
limited number of nuclear charges for which results are presented. In the present work we use a
more complete tabulation from Ref.~\cite{artemyev:05:pra}, originally calculated for He-like
ions. We convert these results from He-like ions to Li-like ions, using the fact that the
following exact relation exists between the $1/Z$ screening QED corrections for Li-like and
He-like ions (see Eq.~(70) of Ref.~\cite{artemyev:05:pra}),
\begin{align} \label{eq:he2li}
E_{(1s)^22p_{1/2}} = \frac12\,E_{(1s\,2p_{1/2})_0} + \frac32\,E_{(1s\,2p_{1/2})_1}\,,\\
E_{(1s)^22p_{3/2}} = \frac34\,E_{(1s\,2p_{3/2})_1} + \frac54\,E_{(1s\,2p_{3/2})_2}\,.
\label{eq:he2lia}
\end{align}

Specifically, for nuclear charges $Z\ge20$, we interpolate the numerical data presented in
Ref.~\cite{artemyev:05:pra}. Values for $Z < 20$ were obtained by fitting numerical data for
$Z\ge20$ to the $\Za$-expansion form
\begin{align}
E_{\rm QEDscr1} = &\ \alpha^2 (\Za)^3\,\Big[D_{1,5}
  \nonumber \\ &
+ (\Za)^2\ln(\Za)\,D^{\rm log}_{1,7} + (\Za)^2 \,D_{1,7} + \ldots\Big]\,,
\end{align}
using the accurate value for the leading coefficient $D_{1,5} = -0.065\,060$, obtained in
Sec.~\ref{sec:NRQED} from fitting the NRQED results for the ${\cal E}^{(5,0)}$ correction.
Numerical results for $E_{\rm QEDscr1}$ are listed in the column ``$1/Z$'' of
Table~\ref{tab:qedscr}.

The column ``$1/Z^2$'' of Table~\ref{tab:qedscr} presents numerical results for the second-order
$1/Z^2$ screening QED correction, $E_{\rm QEDscr2}$, obtained by the amm$+$MQED approach
described in Sec.~\ref{sec:radqed}. The Dirac spectrum is represented by using the DKB method
\cite{shabaev:04:DKB} with $N = 85$ $B$-spline basis functions. The angular integration in radial
matrix elements of the amm operators was carried out according to formulas presented in
Appendix~\ref{app:amm}.  The $\Za$ expansion of $E_{\rm QEDscr2}$ is
\begin{align}
E_{\rm QEDscr2} = &\ \alpha^3 (\Za)^2\,D_{2,5} + \ldots\,,
\end{align}
where $D_{2,5} = 0.1377$ is obtained in Sec.~\ref{sec:NRQED} from fitting the variational NRQED
results for the ${\cal E}^{(5,0)}$ correction. The uncertainty ascribed to this correction in
Table~\ref{tab:qedscr} estimates the error of the approximation. It was evaluated by taking the
difference of the amm$+$MQED and full-QED results for the $1/Z$ screening correction, scaling it
by the ratio $D_{2,5}/(Z\,D_{1,5})$, and multiplying it by  a conservative factor of 2.

The higher-order screening QED correction $E_{\rm QEDscr3+}$ was obtained from the NRQED
calculations described in Sec.~\ref{sec:NRQED}.  For $Z \le 13$, we obtain $E_{\rm \ge4phot}$ by
subtracting the first terms of the $1/Z$ expansion from the $m\alpha^5$ NRQED contribution listed
in Table~\ref{tab:NRQED},
\begin{align}\label{eq:qedscr3}
E_{\rm QEDscr3+} = \alpha^5 {\cal E}^{(5,0)} - \alpha(\Za)^4\left[ D_{0,5} + \frac{D_{1,5}}{Z}
 + \frac{D_{2,5}}{Z^2}\right]\,.
\end{align}
For $Z > 13$, we evaluate $E_{\rm QEDscr3+}$ by fitting the $1/Z$ expansion of numerical results
for $E^{(5,0)}$ listed in Table~\ref{tab:NRQED}. Our results for $E_{\rm QEDscr3+}$ are listed in
Table~\ref{tab:struct}, in the column labeled ``$1/Z^{3+}$''. The indicated numerical uncertainty
takes into account uncalculated QED effects. We estimate these effects by taking the relative
value of the deviation of the full QED results for the $1/Z$ screening correction from the
$m\alpha^5$ contribution induced by the coefficient $D_{1,5}$, and multiplying it by a
conservative factor of 2.

\subsection{Nuclear recoil}

The one-electron nuclear recoil correction $E_{\rm rec}^{\rm oneel}$ was calculated rigorously
within QED to all orders in $\Za$ in Refs.~\cite{artemyev:95:pra,artemyev:95:jpb}. In this work
we take numerical results for $E_{\rm rec}^{\rm oneel}$ from the recent tabulation presented in
Ref.~\cite{yerokhin:15:Hlike}.

The few-body recoil correction $E_{\rm rec}^{\rm fewel}$ is obtained from the NRQED calculations
described in Sec.~\ref{sec:NRQED}.  Specifically, we calculate $E_{\rm rec}^{\rm fewel}$ from
${\cal E}^{(4,1)}$ as
\begin{align}
E_{\rm rec}^{\rm fewel}  =
\alpha^4\, \frac{m}{M}\,  \left(
  {\cal E}^{(4,1)} + \frac{Z^4}{32}\right)\,,
\end{align}
where the second term in braces subtracts the one-electron contribution already taken into
account by $E_{\rm rec}^{\rm oneel}$. For $Z\le 13$, we use the values of ${\cal E}^{(4,1)}$
listed in Table~\ref{tab:NRQED}, whereas for larger $Z$, we get results by fitting the $1/Z$
expansion of ${\cal E}^{(4,1)}$. The uncertainty of the few-body recoil contribution was
estimated by taking the relative value of the deviation of the one-electron QED recoil correction
from the leading-order $m\alpha^4$ term and multiplying it by a conservative factor of 2.

\subsection{Total fine structure}

Table~\ref{tab:theo} summarizes results of our calculations of the $2p_{3/2}\,$--$\,2p_{1/2}$
fine-structure interval in Li-like ions with nuclear chargers $Z = 5\,$--$\,92$. The column
labeled ``$\lbr r^2\rbr^{1/2}$'' contains values for the root-mean-square nuclear charges radii
used in the calculation, taken from Ref.~\cite{angeli:13}. The next column specifies the isotope
for which the calculation is performed. The nuclear masses were taken from Ref.~\cite{wang:12}.

The next three columns display the theoretical results for the electron-structure contribution,
the one-electron QED effects, and the recoil correction, respectively. The one-electron QED part
was taken from the tabulation \cite{yerokhin:15:Hlike}; the other contributions are evaluated as
described in previous Sections.

Results collected in Table~\ref{tab:theo} indicate that for light ions, the dominant theoretical
uncertainty comes from the electron-structure effects, more specifically, from the numerical
uncertainty of the two-photon QED correction and the residual three-photon QED effects. In the
high-$Z$ region, comparable uncertainties arise from various contributions, including the
one-electron QED effects, QED screening, and nuclear charge radii.

\section{Discussion}

Table~\ref{tab:compar} presents a comparison of our theoretical predictions with previous
theoretical and experimental results. For $Z\le 10$, we compare our results with theoretical
values by Wang {\em et al.}~\cite{wang:93}. Their calculation accounted for the
electron-correlation effects within the Breit-Pauli approximation and added the relativistic and
QED effects as delivered by the hydrogenic approximation with an effective nuclear charge. Their
approach is reasonably adequate for very low $Z$. As $Z$ increases, we observe a steadily growing
deviation between their values and our results.

For $Z\ge 10$, we compare our results with the two most complete {\em ab initio} QED
calculations, by Kozhedub {\em et al.}~\cite{kozhedub:10} and by Sapirstein and Cheng
\cite{sapirstein:11}. In these studies, results were reported for the $2p_{3/2}$--$2s$ and
$2p_{1/2}$--$2s$ transition energies; we combine them together to get results for the
$2p_{3/2}$--$2p_{1/2}$ interval. Doing this, we assume the uncertainties of the two transitions
to be correlated. Specifically, we take the largest of the uncertainties reported for the two
intervals, rather than adding them quadratically.

The calculations by Kozhedub {\em et al.} and by Sapirstein and Cheng provided accurate
theoretical predictions for medium- and high-$Z$ ions. For lower-$Z$ ions, however, the relative
accuracy of their results diminished, due to a large cancelation of various effects between the
$2p_{3/2}$ and the $2p_{1/2}$ states. We observe very good agreement with predictions by Kozhedub
{\em et al.} for all nuclear charges reported in that work, well within their error bars. The
agreement with the calculation by Sapirstein and Cheng is good for high values of $Z$ but
moderate in the interval $Z=20\,$--$\,30$, which might be due to residual electron-correlation
effects not accounted for in their work. Our results are significantly more accurate than those
of the both previous studies, partly due to a more complete inclusion of many-photon
electron-correlation effects and partly due to usage of the advantages offered by the
$2p_{3/2}\,$--$\,2p_{1/2}$ interval as compared to the $2p_j\,$--$\,2s$ intervals.

The comparison of our theoretical predictions with the available experimental results is
summarized in Table~\ref{tab:compar} and shows good agreement in most cases. In several occasions
(notably, for $Z = 15$, 39, 82) deviations of about two experimental uncertainties are observed.
The reasons behind them are probably on the experimental side, since
different calculations agree well with each other on the level of the
experimental uncertainties.

Generally, the theoretical predictions for the $2p_{3/2}\,$--$\,2p_{1/2}$ interval are found to
be more accurate than the existing experimental results. The only exception in the range of
nuclear charges covered in this work is boron ($Z = 5$), where the uncertainty of the
experimental result \cite{litzen:98} matches the theoretical accuracy. Even more accurate
measurements are available for Li and Be$^+$ \cite{brown:13,nortenhauser:15}. Unfortunately, our
present approach is not useful for these lightest atoms, since it is relies on the $1/Z$
expansion for description of QED effects of order $m\alpha^6$ and higher, which fails at very low
$Z$.

In summary, we performed {\em ab initio} QED calculations of the $2p$ fine-structure interval in
Li-like ions with nuclear charges $Z = 5\,$--$\,92$. In order to improve the theoretical
accuracy, we combined together two complementary theoretical methods, namely, the $1/Z$-expansion
approach, which accounts for all orders in the parameter $Z\alpha$ but expands in $1/Z$, and the
NRQED approach, which accounts for all orders in $1/Z$ but expands in $\Za$. In the result, we
obtain the currently most accurate theoretical predictions for a wide range of nuclear charges.
For $Z \ge 20$, our theoretical predictions have the fractional accuracy of better than
$10^{-5}$, providing an opportunity for high-precision tests of the interplay of QED and
electron-correlation effects.

\begin{acknowledgments}
The authors are grateful to A.~Kramida for many valuable communications. Work presented in this
paper was supported by the Russian Science Foundation (Grant No. 20-62-46006). K.~P. acknowledges
support from the National Science Center (Poland) (Grant No. 2017/27/B/ST2/02459).
\end{acknowledgments}

\begin{table*}
\caption{The electron structure corrections to the $2p_{3/2}$--$2p_{1/2}$ fine structure splitting, in eV.
\label{tab:struct}}
\begin{ruledtabular}
\begin{tabular}{lw{6.10}w{3.7}w{3.7}w{2.9}w{4.9}w{4.9}w{4.9}w{4.9}}
 $Z$
    & \multicolumn{1}{c}{Dirac }
      & \multicolumn{1}{c}{1-ph. }
          & \multicolumn{2}{c}{2-ph. }
                  & \multicolumn{1}{c}{3-ph. }
                      & \multicolumn{1}{c}{$\ge$4-ph. }
           & \multicolumn{1}{c}{Sum}
               & \multicolumn{1}{c}{Other}
 \\
 & &
          & \multicolumn{1}{c}{MBPT}
          & \multicolumn{1}{c}{QED}
 & & &    & \multicolumn{1}{c}{methods}
 \\\hline\\[-5pt]
%
  5  &  0.028\,325         &   -0.039\,553     &      0.018\,070    &        0.000\,002\,(1)   &    -0.002\,681\,(1)   &   0.000\,050       &     0.004\,214\,(2)   & 0.004\,215^a \\
  6  &  0.058\,757         &   -0.068\,384     &      0.026\,043    &        0.000\,004\,(2)   &    -0.003\,224\,(2)   &   0.000\,057       &     0.013\,253\,(3)   & 0.013\,249^a \\
  7  &  0.108\,901         &   -0.108\,658     &      0.035\,483    &        0.000\,007\,(3)   &    -0.003\,771\,(3)   &   0.000\,062       &     0.032\,023\,(5)   & 0.031\,998^a \\
  8  &  0.185\,873         &   -0.162\,311     &      0.046\,398    &        0.000\,012\,(5)   &    -0.004\,323\,(5)   &   0.000\,065\,(1)  &     0.065\,714\,(7)   & 0.065\,63^a \\
  9  &  0.297\,902         &   -0.231\,290     &      0.058\,798    &        0.000\,018\,(7)   &    -0.004\,880\,(6)   &   0.000\,067\,(1)  &     0.120\,615\,(9)   & 0.120\,40^a \\
 10  &  0.454\,338         &   -0.317\,558     &      0.072\,695    &        0.000\,027\,(9)   &    -0.005\,444\,(8)   &   0.000\,069\,(1)  &     0.204\,128\,(12)  & 0.203\,66^a \\
     &                     &                      &                    &                         &                      &                     &                     & 0.204\,1\,(6)^b\\
 15  &  2.309\,735         &   -1.078\,183     &      0.165\,127    &        0.000\,128\,(3)   &    -0.008\,373\,(26)  &   0.000\,074\,(3)  &     1.388\,51\,(3)    & 1.388\,4\,(3)^b \\
 20  &  7.343\,045         &   -2.577\,266     &      0.297\,502    &        0.000\,347\,(6)   &    -0.011\,556\,(54)  &   0.000\,076\,(5)  &     5.052\,15\,(5)    & 5.052\,4\,(3)^b \\
 26  &  21.169\,98         &   -5.738\,39      &      0.513\,50     &        0.000\,94\,(1)    &    -0.015\,84\,(12)   &   0.000\,08\,(1)   &     15.930\,26\,(12)  & 15.930\,9\,(3)^b \\
 28  &  28.580\,10         &   -7.204\,91      &      0.600\,50     &        0.001\,24\,(1)    &    -0.017\,41\,(14)   &   0.000\,08\,(1)   &     21.959\,59\,(14)  & 21.960\,5\,(3)^b \\
 30  &  37.813\,57         &   -8.912\,13      &      0.695\,52     &        0.001\,62\,(1)    &    -0.019\,07\,(18)   &   0.000\,08\,(1)   &     29.579\,58\,(18)  & 29.579\,6\,(3)^b \\
 36  &  79.495\,27         &   -15.703\,91     &      1.032\,61     &        0.003\,19\,(2)    &    -0.024\,61\,(30)   &   0.000\,08\,(2)   &     64.802\,6\,(3)    & 64.803\,3\,(5)^b \\
 40  &  122.466\,49        &   -21.871\,21     &      1.305\,26     &        0.004\,73\,(3)    &    -0.028\,87\,(42)   &   0.000\,08\,(2)   &     101.876\,5\,(4)   & 101.878\,4\,(11)^b \\
 47  &  238.588\,48\,(1)   &   -36.594\,23     &      1.890\,10     &        0.008\,81\,(4)    &    -0.037\,70\,(70)   &   0.000\,08\,(3)   &     203.855\,5\,(7)   & 203.856\,6\,(16)^b \\
 50  &  308.853\,99\,(1)   &   -44.725\,25     &      2.188\,88     &        0.011\,19\,(4)    &    -0.042\,12\,(86)   &   0.000\,08\,(4)   &     266.286\,8\,(9)   & 266.288\,1\,(21)^b \\
 54  &  426.722\,24\,(3)   &   -57.580\,77     &      2.639\,24     &        0.015\,11\,(5)    &    -0.048\,7\,(11)    &   0.000\,08\,(4)   &     371.747\,2\,(11)  & 371.748\,7\,(29)^b \\
 60  &  667.503\,4\,(1)    &   -81.929\,1      &      3.443\,8      &        0.022\,9\,(1)     &    -0.060\,5\,(16)    &   0.000\,1\,(1)    &     588.980\,6\,(16)  & 588.983\,4\,(41)^b \\
 70  &  1302.156\,8\,(4)   &   -139.881\,9     &      5.224\,0      &        0.042\,1\,(1)     &    -0.086\,5\,(28)    &   0.000\,1\,(1)    &     1167.455\,(3)     & 1167.461\,(11)^b \\
 80  &  2367.736\,6\,(16)  &   -228.290\,4     &      7.783\,3      &        0.071\,4\,(2)     &    -0.124\,4\,(46)    &   0.000\,1\,(1)    &     2147.177\,(5)     & 2147.188\,(14)^b \\
 83  &  2804.106\,0\,(23)  &   -262.859\,7     &      8.761\,3      &        0.082\,3\,(2)     &    -0.139\,0\,(52)    &   0.000\,1\,(1)    &     2549.951\,(6)     & 2549.961\,(16)^b \\
 90  &  4103.324\,(12)     &   -362.755\,8     &      11.560\,8     &        0.110\,8\,(2)     &    -0.181\,7\,(70)    &   0.000\,1\,(2)    &     3752.058\,(14)    & 3752.127\,(41)^b \\
 92  &  4561.237\,4\,(47)  &   -397.245\,0     &      12.523\,2     &        0.119\,8\,(3)     &    -0.196\,6\,(75)    &   0.000\,1\,(2)    &     4176.439\,(9)     & 4176.457\,(51)^b \\
\end{tabular}
\end{ruledtabular}
$^a$ NRQED, this work; $^b$
Kozhedub {\em et al.} \cite{kozhedub:10}.
\end{table*}

\begin{table*}
\caption{The screened QED corrections to the $2p_{3/2}$--$2p_{1/2}$ fine structure splitting, in eV.
\label{tab:qedscr}}
\begin{ruledtabular}
\begin{tabular}{lw{4.12}w{4.12}w{4.12}w{4.12}w{4.10}w{4.10}}
 $Z$ & \multicolumn{1}{c}{$1/Z^1$ }
       & \multicolumn{1}{c}{$1/Z^{2}$ }
       & \multicolumn{1}{c}{$1/Z^{3+}$ }
           & \multicolumn{1}{c}{Sum}
               & \multicolumn{1}{c}{NRQED}
               & \multicolumn{1}{c}{Kozhedub {\em et al.} \cite{kozhedub:10}}
 \\\hline\\[-5pt]
  5    &   -0.000\,084\,6\,(1)   &   0.000\,035\,6\,(2)    &   -0.000\,004\,8\,(2)   &   -0.000\,053\,8\,(3)  & -0.000\,054         \\
  6    &   -0.000\,145\,4\,(2)   &   0.000\,050\,9\,(5)    &   -0.000\,005\,8\,(3)   &   -0.000\,100\,3\,(6)  & -0.000\,102         \\
  7    &   -0.000\,229\,5\,(4)   &   0.000\,068\,6\,(8)    &   -0.000\,006\,8\,(4)   &   -0.000\,168\,(1)   & -0.000\,171         \\
  8    &   -0.000\,340\,3\,(7)   &   0.000\,088\,7\,(13)   &   -0.000\,007\,8\,(5)   &   -0.000\,259\,(2)   & -0.000\,267         \\
  9    &   -0.000\,481\,1\,(12)  &   0.000\,111\,2\,(20)   &   -0.000\,008\,8\,(7)   &   -0.000\,379\,(2)   & -0.000\,392         \\
 10    &   -0.000\,655\,1\,(20)  &   0.000\,136\,0\,(29)   &   -0.000\,009\,8\,(9)   &   -0.000\,529\,(4)   & -0.000\,552  & -0.000\,5\,(2)  \\
 12    &   -0.001\,114\,2\,(43)  &   0.000\,191\,6\,(54)   &   -0.000\,011\,8\,(15)  &   -0.000\,935\,(7)    & -0.000\,991  & -0.000\,9\,(3)  \\
 15    &   -0.002\,120\,(11)     &   0.000\,289\,(12)      &   -0.000\,015\,(3)      &   -0.001\,85\,(2)    &              & -0.001\,8\,(4)  \\
 18    &   -0.003\,558\,(22)     &   0.000\,399\,(22)      &   -0.000\,018\,(4)      &   -0.003\,18\,(3)    &              & -0.003\,2\,(5)  \\
 20    &   -0.004\,790\,(27)     &   0.000\,478\,(29)      &   -0.000\,020\,(5)      &   -0.004\,33\,(4)    &              & -0.004\,3\,(5)  \\
 26    &   -0.009\,83\,(15)      &   0.000\,731\,(69)      &   -0.000\,026\,(10)     &   -0.009\,1\,(2)     &              & -0.009\,2\,(8)  \\
 30    &   -0.014\,30\,(8)       &   0.000\,90\,(11)       &   -0.000\,030\,(14)     &   -0.013\,4\,(2)     &              & -0.013\,6\,(11)  \\
 32    &   -0.016\,85\,(13)      &   0.000\,98\,(14)       &   -0.000\,032\,(16)     &   -0.015\,9\,(2)     &              & -0.016\,0\,(12)  \\
 40    &   -0.028\,56\,(14)      &   0.001\,22\,(31)       &   -0.000\,040\,(28)     &   -0.027\,4\,(4)     &              & -0.027\,9\,(18)  \\
 47    &   -0.039\,21\,(26)      &   0.001\,25\,(55)       &   -0.000\,047\,(42)     &   -0.038\,0\,(7)     &              & -0.038\,7\,(24)  \\
 50    &   -0.043\,02\,(29)      &   0.001\,18\,(69)       &   -0.000\,050\,(50)     &   -0.041\,9\,(9)     &              & -0.042\,8\,(27)  \\
 54    &   -0.046\,78\,(38)      &   0.000\,99\,(89)       &   -0.000\,054\,(61)     &   -0.045\,9\,(10)    &              & -0.047\,0\,(32)  \\
 60    &   -0.046\,21\,(45)      &   0.000\,5\,(13)        &   -0.000\,060\,(82)     &   -0.045\,8\,(14)    &              & -0.048\,0\,(42)  \\
 66    &   -0.034\,39\,(64)      &   -0.000\,4\,(19)       &   -0.000\,07\,(11)      &   -0.035\,(2)      &              & -0.037\,(5)  \\
 70    &   -0.014\,83\,(47)      &   -0.001\,2\,(24)       &   -0.000\,07\,(13)      &   -0.016\,(2)      &              & -0.020\,(7)  \\
 74    &    0.015\,98\,(86)      &   -0.002\,1\,(29)       &   -0.000\,07\,(16)      &    0.014\,(3)      &              &  0.010\,(8)  \\
 80    &    0.098\,74\,(58)      &   -0.003\,3\,(40)       &   -0.000\,08\,(20)      &    0.095\,(4)      &              &  0.086\,(11) \\
 82    &    0.138\,13\,(63)      &   -0.003\,5\,(44)       &   -0.000\,08\,(22)      &    0.135\,(4)      &              &  0.122\,(12)  \\
 90    &    0.386\,16\,(82)      &   -0.002\,0\,(66)       &   -0.000\,09\,(32)      &    0.384\,(7)      &              &  0.359\,(17)  \\
 92    &    0.478\,80\,(91)      &   -0.000\,3\,(73)       &   -0.000\,09\,(35)      &    0.478\,(7)      &              &  0.446\,(19)  \\
\end{tabular}
\end{ruledtabular}
\end{table*}

\begin{table*}
\caption{Comparison of different theoretical predictions and experimental results for the $2p_{3/2}$--$2p_{1/2}$ fine-structure interval
in Li-like ions, in cm$^{-1}$ or eV as indicated, 1~eV = $8065.543\,937\,$~cm$^{-1}$.
\label{tab:compar}}
\begin{ruledtabular}
\begin{tabular}{lw{4.10}....c}
 $Z$ & \multicolumn{1}{c}{This work }
       & \multicolumn{1}{c}{Wang 1993 \cite{wang:93}}
       & \multicolumn{1}{c}{Kozhedub 2010 \cite{kozhedub:10}}
       & \multicolumn{1}{c}{Sapirstein 2011 \cite{sapirstein:11}}
       & \multicolumn{1}{c}{Experiment}
           & \multicolumn{1}{c}{Ref.}
 \\\hline\\[-5pt]
\multicolumn{1}{l}{in cm$^{-1}$:}\\
  5     & 34.075\,(13)      &  34x.04    &               &              & 34x.100\,(14) & \cite{litzen:98} \\
  6     & 107.166\,(23)     & 107x.06    &               &              & 107x.3\,(3)   & \cite{boyce:35,edlen:83} \\
  7     & 258.931\,(37)     & 258x.7     &               &              & 259x\,(1)     & \cite{edlen:83} \\
  8     & 531.323\,(55)     & 530x.9     &               &              & 531x\,(1)     & \cite{edlen:83} \\
  9     & 975.206\,(77)     & 974x.5     &               &              & 976x\,(2)     & \cite{edlen:83} \\
 10     & 1\,650.39\,(10)   & 1649x.2    & 1653x\,(3)   &  1653x\,(8)   & 1649x\,(2)    & \cite{edlen:83} \\
 11     & 2\,625.73\,(10)   &            &               &              & 2631x\,(5)    & \cite{edlen:83} \\
 12     & 3\,979.15\,(13)   &           &               &  3984x\,(8)   & 3975x\,(3)    & \cite{edlen:83} \\
 13     & 5\,797.76\,(16)   &            &               &              & 5796x\,(5)    & \cite{edlen:83} \\
 14     & 8\,177.95\,(21)   &            &               &              & 8177x\,(4)    & \cite{edlen:83} \\
 15     & 11\,225.38\,(25)  &           & 11224x\,(4)   &  11219x\,(8)  & 11253x\,(15)  & \cite{edlen:83} \\
 16     & 15\,055.24\,(30)  &           &               &               & 15054x\,(1)   & \cite{thomas:94}\\
 17     & 19\,792.36\,(35)  &            &               &              & 19770x\,(15)  & \cite{edlen:83} \\
 18     & 25\,571.24\,(42)  &           & 25572x\,(5)   &  25560x\,(8)  & 25572x\,(10)  & \cite{edlen:83} \\
 20     & 40\,841.36\,(55)  &           & 40843x\,(6)   &  40828x\,(8)  & 40850x\,(10)  & \cite{edlen:83} \\
 21     & 50\,651.62\,(70)  &           &               &  50627x\,(8)  \\
 22     & 62\,141.83\,(95)  &            &               &              & 62146x\,(10)  & \cite{edlen:83} \\
 24     & 90\,914.5\,(15)   &            &               &              & 90912x\,(12)  & \cite{sugar:93}\\
 25     & 108\,598.5\,(16)  &            &               &              & 108634x\,(40) & \cite{edlen:83} \\
 26     & 128\,769.8\,(17)  &            & 128774x\,(7)  & 128750x\,(8) & 128774x\,(16) & \cite{reader:94}\\
 28     & 177\,502.2\,(17)  &            & 177508x\,(8)  & 177474x\,(8) & 177524x\,(20) & \cite{sugar:92} \\
 29     & 206\,557.7\,(17)  &            &               &              & 206549x\,(33) & \cite{knize:91}
\end{tabular}
\end{ruledtabular}
\vskip0.25cm
\begin{ruledtabular}
\begin{tabular}{lw{4.10}w{4.10}w{4.10}w{4.10}c}
 $Z$ & \multicolumn{1}{c}{This work }
       & \multicolumn{1}{c}{Kozhedub 2010 \cite{kozhedub:10}}
       & \multicolumn{1}{c}{Sapirstein 2011 \cite{sapirstein:11}}
       & \multicolumn{1}{c}{Experiment}
           & \multicolumn{1}{c}{Ref.}
 \\\hline\\[-5pt]
\multicolumn{1}{l}{in eV:}\\
 30     & 29.643\,27\,(23)  & 29.643\,6\,(12) & 29.641\,(1)      & 29.646\,4\,(47) & \cite{staude:98}\\
 32     & 39.142\,30\,(30)  &                 & 39.14            & 39.141\,7\,(53) & \cite{knize:91} \\
 34     & 50.799\,46\,(38)  &                 &                  & 50.790\,(23) & \cite{hinnov:89}  \\
 36     & 64.936\,53\,(43)  & 64.936\,7\,(17) & 64.93            & 64.955\,(37) & \cite{madzunkov:02,podpaly:14}\\
 39     & 91.565\,77\,(53)  &                 &                  & 91.595\,(15) & \cite{silwal:17}  \\
 40     & 102.080\,50\,(58) & 102.081\,7\,(23)& 102.08  \\
 42     & 125.879\,40\,(70) &                 & 125.88           & 125.841\,(73) & \cite{hinnov:89} \\
 47     & 204.238\,9\,(11)  & 204.238\,8\,(36)& 204.26           & 204.229\,(31) & \cite{bosselmann:99} \\
 50     & 266.772\,5\,(14)  & 266.772\,1\,(46)& 266.77      \\
 52     & 316.135\,1\,(16)  & 316.134\,(5)    & 316.11\\
 54     & 372.395\,0\,(19)  &                 & 372.39           & 372.354\,(53) & \cite{feili:00,bernhardt:15}\\
 60     & 589.928\,5\,(30)  & 589.929\,(6)    & 589.93\,(1)  \\
 64     & 784.028\,3\,(41)  &                 & 784.01\,(1)    \\
 66     & 898.712\,1\,(48)  &                 & 898.73\,(1)    \\
 70     & 1\,169.031\,3\,(49) &               & 1\,169.03\,(2)   \\
 74     & 1\,502.715\,0\,(65) &               & 1\,502.66\,(3)   \\
 79     & 2\,027.756\,9\,(93) &               & 2\,027.78\,(3)  \\
 80     & 2\,149.404\,(10)    &               & 2\,149.41\,(4)  \\
 82     & 2\,411.403\,(11)    &               & 2\,411.41\,(4) & 2\,411.61\,(12) & \cite{brandau:03,zhang:08} \\
 83     & 2\,552.326\,(11)    &               & 2\,552.32\,(5)  \\
 90     & 3\,754.525\,(22)    &               & 3\,754.51\,(7)  \\
 92     & 4\,178.830\,(22)    &               & 4\,178.81\,(8) & 4\,178.73\,(21) & \cite{beiersdorfer:05,beiersdorfer:93}\\
\end{tabular}
\end{ruledtabular}
\end{table*}

\appendix

\begin{widetext}

\section{Radial integrations in matrix elements of amm operators}
\label{app:amm}

In this section we present formulas for the matrix elements of the amm operators [given by
Eqs.~(\ref{eq:amm1}) and (\ref{eq:amm2})] with the Dirac wave functions, after angular
integrations. The matrix element of the one-electron amm operator is evaluated as
\begin{align}
\lbr a | H_{\rm amm, 1}|b\rbr =  -\frac{\Za\kappa}{4} \,\delta_{\kappa_a\kappa_b}\,\delta_{\mu_a\mu_b}
\, \int_{0}^{\infty}r^2dr\, \frac1{r^2}\,\big[ g_{a}(r)\,f_{b}(r) + f_a(r)\,g_b(r)\big]\,,
\end{align}
where $g_n(r)$ and $f_n(r)$ are the upper and the lower radial components of the wave function of
the electron state $n$, defined as in Ref.~\cite{yerokhin:20:green}; $\kappa_n$ and $\mu_n$ are
the relativistic angular-momentum quantum number and the angular-momentum projection,
correspondingly.

The matrix element of the two-electron amm operator can be written in the form, analogous to that
for the matrix element of the electron-electron interaction operator ({\em cf.} Eq.~(38) in
Ref.~\cite{yerokhin:20:green}),
\begin{align}
\lbr ab | H_{\rm amm, 2}|cd\rbr =  \frac{\alpha\kappa}{4}
  \sum_L J_L(abcd)\, R^{\rm amm, 2}_L(abcd)\,,
\end{align}
where $J_L(abcd)$ is the standard function incorporating the dependence of a two-body operator on
the angular-momentum projections (see Eq.~(39) of Ref.~\cite{yerokhin:20:green}) and $R^{\rm amm,
2}_L$ is the radial integral evaluated as
\begin{align}
R^{\rm amm, 2}_L(abcd) = &\ (-1)^L\,(2L+1)\,\int_{0}^{\infty}r_1^2dr_1\,
 \Bigg[
    \sqrt{\frac{L+1}{2L+1}}\,C_{L}(\kappa_b,\kappa_d)\,
        \frac1{r_1^{L+2}}\,X_{ac,LL+1}(r_1)\, \int_0^{r_1}r_2^2dr_2\, r_2^L\, W_{bd}(r_2)
     \nonumber   \\
 & +
    \sqrt{\frac{L}{2L+1}}\,C_{L}(\kappa_b,\kappa_d)\,
        \frac1{r_1^{L+1}}\,W_{bd}(r_1)\, \int_0^{r_1}r_2^2dr_2\, r_2^{L-1}\,X_{ac,LL-1}(r_2)
     \nonumber   \\
 & +
  \sum_{l = L-1}^L  \sqrt{6(l+1)}\,
           \SixJ{1}{1}{1}{L}{l}{l+1}
        \frac1{r_1^{l+2}}\,Y_{ac,Ll+1}(r_1)\, \int_0^{r_1}r_2^2dr_2\, r_2^{l}\,Z_{bd,Ll}(r_2)
     \nonumber   \\
 & +
  \sum_{l = L}^{L+1}  \sqrt{6l}\,
           \SixJ{1}{1}{1}{L}{l}{l-1}
        \frac1{r_1^{l+1}}\,Z_{bd,Ll}(r_1)\, \int_0^{r_1}r_2^2dr_2\, r_2^{l-1}\,Y_{ac,Ll-1}(r_2)
+ \ldots (ac) \leftrightarrow (bd) \ldots \Bigg]\,,
\end{align}
where $\{\ldots\}$ denotes the 6$j$-symbol and
\begin{align}
X_{ac,ll'}(r) = & \ g_a(r)\,f_c(r)\,S_{ll'}(-\kappa_c,\kappa_a) + f_a(r)\,g_c(r)\,S_{ll'}(\kappa_c,-\kappa_a)
 \,, \\
Y_{ac,ll'}(r) = & \ g_a(r)\,g_c(r)\,S_{ll'}(\kappa_c,\kappa_a) - f_a(r)\,f_c(r)\,S_{ll'}(-\kappa_c,-\kappa_a)
 \,, \\
Z_{ac,ll'}(r) = & \ g_a(r)\,f_c(r)\,S_{ll'}(-\kappa_c,\kappa_a) - f_a(r)\,g_c(r)\,S_{ll'}(\kappa_c,-\kappa_a)
 \,, \\
W_{ac}(r) = & \ g_a(r)\,g_c(r) + f_a(r)\,f_c(r)
 \,.
\end{align}
Furthermore, the standard angular coefficients $S_{ll'}$ and $C_l$ are defined by Eqs.~(A7)-(A10)
of Ref.~\cite{yerokhin:20:green}.

\end{widetext}


\begin{longtable*}{lcw{4.9}w{5.10}w{4.10}w{4.11}w{4.10}w{4.10}}
\caption{Individual effects and total theoretical predictions for the $2p_{3/2}$--$2p_{1/2}$
fine-structure interval in Li-like ions. Units are eV, 1~eV = $8065.543\,937\,$~cm$^{-1}$. In the
case when an entry is given with two uncertainties, the first one is the estimation of the
theoretical error and the second is due to the nuclear charge radius. In the case when one
uncertainty is given, it is the estimation of the theoretical error and the uncertainty due to
the nuclear radius is negligible. \label{tab:theo}}
\\ \colrule\hline\\[-7pt]
 $Z$ &  Isotope & \multicolumn{1}{c}{$\lbr r^2\rbr^{1/2}$ [fm]}
       & \multicolumn{1}{c}{Structure}
       & \multicolumn{1}{c}{QED,1-el}
       & \multicolumn{1}{c}{QED,scr}
       & \multicolumn{1}{c}{Recoil}
       & \multicolumn{1}{c}{Total}
 \\\hline\\[-5pt]
%
  5     &  $^{ 11}$B  & 2.406\,(29)  &   0.004\,213\,7\,(16)    & 0.000\,065\,2          & -0.000\,053\,8\,(3)    & -0.000\,0003           & 0.004\,224\,8\,(17)      \\
  6     &  $^{ 12}$C  & 2.4702\,(22) &   0.013\,253\,2\,(28)    & 0.000\,135\,0          & -0.000\,100\,3\,(6)    & -0.000\,0009           & 0.013\,286\,9\,(29)      \\
  7     &  $^{ 14}$N  & 2.5582\,(70) &   0.032\,023\,3\,(45)    & 0.000\,249\,5          & -0.000\,167\,7\,(10)   & -0.000\,0018           & 0.032\,103\,4\,(46)      \\
  8     &  $^{ 16}$0  & 2.6991\,(52) &   0.065\,713\,6\,(67)    & 0.000\,424\,6\,(1)     & -0.000\,259\,4\,(16)   & -0.000\,0031           & 0.065\,875\,7\,(68)      \\
  9     &  $^{ 19}$F  & 2.8976\,(25) &   0.120\,615\,2\,(93)    & 0.000\,678\,4\,(1)     & -0.000\,378\,7\,(24)   & -0.000\,0048           & 0.120\,910\,1\,(96)      \\
 10     &  $^{ 20}$Ne & 3.0055\,(21) &   0.204\,128\,(12)       & 0.001\,031\,1\,(1)     & -0.000\,529\,0\,(36)   & -0.000\,0076           & 0.204\,622\,(13)       \\
 11     &  $^{ 23}$Na & 2.9936\,(21) &   0.324\,768\,(11)       & 0.001\,505             & -0.000\,713\,(5)       & -0.000\,010            & 0.325\,549\,(12)       \\
 12     &  $^{ 24}$Mg & 3.0570\,(16) &   0.492\,176\,(14)       & 0.002\,126             & -0.000\,934\,(7)       & -0.000\,015            & 0.493\,351\,(16)       \\
 13     &  $^{ 27}$Al & 3.0610\,(31) &   0.717\,127\,(18)       & 0.002\,919             & -0.001\,195\,(9)       & -0.000\,020            & 0.718\,831\,(20)       \\
 14     &  $^{ 28}$Si & 3.1224\,(24) &   1.011\,548\,(22)       & 0.003\,913\,(1)        & -0.001\,498\,(12)      & -0.000\,027            & 1.013\,936\,(26)       \\
 15     &  $^{ 31}$P  & 3.1889\,(19) &   1.388\,508\,(26)       & 0.005\,140\,(1)        & -0.001\,846\,(16)      & -0.000\,033            & 1.391\,770\,(31)       \\
 16     &  $^{ 32}$S  & 3.2611\,(18) &   1.862\,263\,(31)       & 0.006\,632\,(1)        & -0.002\,240\,(20)      & -0.000\,043            & 1.866\,612\,(37)       \\
 17     &  $^{ 35}$Cl & 3.365\,(19)  &   2.448\,253\,(36)       & 0.008\,422\,(2)        & -0.002\,683\,(25)      & -0.000\,052            & 2.453\,940\,(44)       \\
 18     &  $^{ 40}$Ar & 3.4274\,(26) &   3.163\,116\,(41)       & 0.010\,548\,(3)        & -0.003\,177\,(31)      & -0.000\,058            & 3.170\,429\,(52)       \\
 19     &  $^{ 39}$K  & 3.4349\,(19) &   4.024\,704\,(46)       & 0.013\,047\,(4)        & -0.003\,723\,(38)      & -0.000\,076            & 4.033\,952\,(59)       \\
 20     &  $^{ 40}$Ca & 3.4776\,(19) &   5.052\,148\,(54)       & 0.015\,959\,(5)        & -0.004\,331\,(40)      & -0.000\,093            & 5.063\,683\,(68)       \\
 21     &  $^{ 45}$Sc & 3.5459\,(25) &   6.265\,777\,(63)       & 0.019\,324\,(6)        & -0.004\,998\,(59)      & -0.000\,103            & 6.280\,001\,(87)       \\
 22     &  $^{ 48}$Ti & 3.5921\,(17) &   7.687\,235\,(72)       & 0.023\,186\,(9)        & -0.005\,697\,(92)      & -0.000\,118            & 7.704\,60\,(12)        \\
 23     &  $^{ 51}$V  & 3.6002\,(22) &   9.339\,469\,(83)       & 0.027\,587\,(11)       & -0.006\,48\,(12)       & -0.000\,135            & 9.360\,44\,(15)        \\
 24     &  $^{ 52}$Cr & 3.6452\,(42) &   11.246\,84\,(11)       & 0.032\,574\,(14)       & -0.007\,30\,(15)       & -0.000\,160            & 11.271\,96\,(18)       \\
 25     &  $^{ 55}$Mn & 3.7057\,(22) &   13.434\,66\,(10)       & 0.038\,194\,(18)       & -0.008\,18\,(16)       & -0.000\,180            & 13.464\,50\,(19)       \\
 26     &  $^{ 56}$Fe & 3.7377\,(16) &   15.930\,26\,(12)       & 0.044\,492\,(23)       & -0.009\,12\,(17)       & -0.000\,210            & 15.965\,42\,(21)       \\
 27     &  $^{ 59}$Co & 3.7875\,(21) &   18.761\,95\,(13)       & 0.051\,517\,(29)       & -0.010\,12\,(16)       & -0.000\,235            & 18.803\,11\,(21)       \\
 28     &  $^{ 58}$Ni & 3.7757\,(20) &   21.959\,59\,(14)       & 0.059\,320\,(36)       & -0.011\,17\,(15)       & -0.000\,280            & 22.007\,46\,(21)       \\
 29     &  $^{ 63}$Cu & 3.8823\,(15) &   25.554\,52\,(16)       & 0.067\,951\,(44)       & -0.012\,29\,(14)       & -0.000\,300\,(1)       & 25.609\,89\,(22)       \\
 30     &  $^{ 64}$Zn & 3.9283\,(15) &   29.579\,58\,(18)       & 0.077\,46\,(5)         & -0.013\,43\,(14)       & -0.000\,34             & 29.643\,27\,(23)       \\
 31     &  $^{ 69}$Ga & 3.9973\,(17) &   34.069\,17\,(20)       & 0.087\,90\,(7)         & -0.014\,64\,(16)       & -0.000\,37             & 34.142\,06\,(26)       \\
 32     &  $^{ 74}$Ge & 4.0742\,(12) &   39.059\,27\,(21)       & 0.099\,32\,(8)         & -0.015\,90\,(19)       & -0.000\,39             & 39.142\,30\,(30)       \\
 33     &  $^{ 75}$As & 4.0968\,(20) &   44.587\,54\,(24)       & 0.111\,78\,(10)        & -0.017\,24\,(22)       & -0.000\,44             & 44.681\,63\,(34)       \\
 34     &  $^{ 80}$Se & 4.1400\,(18) &   50.693\,19\,(26)       & 0.125\,32\,(12)        & -0.018\,57\,(25)       & -0.000\,47             & 50.799\,46\,(38)       \\
 35     &  $^{ 79}$Br & 4.1629\,(21) &   57.417\,02\,(25)       & 0.140\,00\,(9)         & -0.019\,98\,(27)       & -0.000\,54             & 57.536\,49\,(38)       \\
 36     &  $^{ 84}$Kr & 4.1884\,(22) &   64.802\,64\,(31)       & 0.155\,86\,(11)        & -0.021\,40\,(29)       & -0.000\,58\,(1)        & 64.936\,53\,(43)       \\
 37     &  $^{ 85}$Rb & 4.2036\,(24) &   72.893\,76\,(33)       & 0.172\,97\,(13)        & -0.022\,87\,(30)       & -0.000\,64\,(1)        & 73.043\,22\,(46)       \\
 38     &  $^{ 88}$Sr & 4.2240\,(18) &   81.737\,18\,(36)       & 0.191\,38\,(15)        & -0.024\,36\,(31)       & -0.000\,69\,(1)        & 81.903\,50\,(49)       \\
 39     &  $^{ 89}$Y  & 4.2430\,(21) &   91.381\,29\,(39)       & 0.211\,12\,(18)        & -0.025\,87\,(32)       & -0.000\,77\,(1)        & 91.565\,77\,(53)       \\
 40     &  $^{ 90}$Zr & 4.2694\,(10) &   101.876\,49\,(42)      & 0.232\,25\,(21)        & -0.027\,39\,(34)       & -0.000\,85\,(1)        & 102.080\,50\,(58)      \\
 41     &  $^{ 93}$Nb & 4.3240\,(17) &   113.275\,21\,(45)      & 0.254\,82\,(24)        & -0.028\,97\,(37)       & -0.000\,91\,(2)        & 113.500\,14\,(63)      \\
 42     &  $^{ 98}$Mo & 4.4091\,(18) &   125.632\,01\,(49)      & 0.278\,87\,(28)        & -0.030\,52\,(41)       & -0.000\,96\,(2)        & 125.879\,40\,(70)      \\
 43     &  $^{ 98}$Tc & 4.424\,(44)  &   139.003\,39\,(50)(4)   & 0.304\,45\,(32)        & -0.032\,07\,(45)       & -0.001\,07\,(2)        & 139.274\,71\,(74)(4)   \\
 44     &  $^{102}$Ru & 4.4809\,(18) &   153.449\,50\,(58)      & 0.331\,60\,(36)        & -0.033\,60\,(49)       & -0.001\,14\,(3)        & 153.746\,36\,(84)      \\
 45     &  $^{103}$Rh & 4.4945\,(23) &   169.030\,26\,(61)      & 0.360\,35\,(42)        & -0.035\,12\,(53)       & -0.001\,24\,(3)        & 169.354\,24\,(91)      \\
 46     &  $^{106}$Pd & 4.5318\,(29) &   185.810\,18\,(66)(1)   & 0.390\,74\,(48)        & -0.036\,59\,(57)       & -0.001\,33\,(4)        & 186.163\,00\,(99)(1)   \\
 47     &  $^{107}$Ag & 4.5454\,(31) &   203.855\,55\,(70)(1)   & 0.422\,80\,(54)        & -0.038\,01\,(61)       & -0.001\,45\,(4)        & 204.238\,9\,(11)       \\
 48     &  $^{112}$Cd & 4.5944\,(24) &   223.235\,19\,(75)(1)   & 0.456\,56\,(62)        & -0.039\,38\,(65)       & -0.001\,52\,(5)        & 223.650\,9\,(12)       \\
 49     &  $^{115}$In & 4.6156\,(26) &   244.020\,82\,(81)(1)   & 0.492\,05\,(70)        & -0.040\,67\,(69)       & -0.001\,62\,(6)        & 244.470\,6\,(13)       \\
 50     &  $^{120}$Sn & 4.6519\,(21) &   266.286\,79\,(86)(1)   & 0.529\,28\,(79)        & -0.041\,89\,(75)       & -0.001\,70\,(6)        & 266.772\,5\,(14)       \\
 51     &  $^{121}$Sb & 4.6802\,(26) &   290.110\,47\,(92)(1)   & 0.568\,27\,(89)        & -0.043\,08\,(80)       & -0.001\,84\,(7)        & 290.633\,8\,(15)       \\
 52     &  $^{130}$Te & 4.7423\,(25) &   315.572\,13\,(98)(2)   & 0.609\,0\,(10)         & -0.044\,14\,(85)       & -0.001\,87\,(8)        & 316.135\,1\,(16)       \\
 53     &  $^{127}$I  & 4.7500\,(81) &   342.755\,5\,(10)       & 0.651\,5\,(11)         & -0.045\,08\,(91)       & -0.002\,08\,(10)       & 343.359\,9\,(18)       \\
 54     &  $^{132}$Xe & 4.7859\,(48) &   371.747\,2\,(11)       & 0.695\,8\,(12)         & -0.045\,85\,(98)       & -0.002\,18\,(11)       & 372.395\,0\,(19)       \\
 55     &  $^{133}$Cs & 4.8041\,(46) &   402.637\,5\,(12)       & 0.741\,8\,(14)         & -0.046\,4\,(10)        & -0.002\,35\,(12)       & 403.330\,5\,(21)       \\
 56     &  $^{138}$Ba & 4.8378\,(46) &   435.520\,0\,(13)       & 0.789\,6\,(16)         & -0.046\,8\,(11)        & -0.002\,45\,(14)       & 436.260\,3\,(23)       \\
 57     &  $^{139}$La & 4.8550\,(49) &   470.492\,2\,(13)(1)    & 0.839\,0\,(17)         & -0.047\,0\,(12)        & -0.002\,64\,(15)       & 471.281\,7\,(25)(1)    \\
 58     &  $^{140}$Ce & 4.8771\,(18) &   507.655\,5\,(14)       & 0.890\,1\,(19)         & -0.046\,9\,(13)        & -0.002\,84\,(18)       & 508.495\,9\,(27)       \\
 59     &  $^{141}$Pr & 4.8919\,(50) &   547.115\,2\,(15)(1)    & 0.942\,8\,(21)         & -0.046\,5\,(13)        & -0.003\,05\,(20)       & 548.008\,4\,(29)(1)    \\
 60     &  $^{142}$Nd & 4.9123\,(25) &   588.980\,6\,(16)(1)    & 0.997\,0\,(21)         & -0.045\,8\,(14)        & -0.003\,30\,(23)       & 589.928\,5\,(30)(1)    \\
 61     &  $^{145}$Pm & 4.962\,(50)  &   633.365\,4\,(17)(8)    & 1.052\,7\,(23)         & -0.045\,1\,(15)        & -0.003\,46\,(25)       & 634.369\,5\,(32)(8)    \\
 62     &  $^{152}$Sm & 5.0819\,(60) &   680.387\,2\,(18)(1)    & 1.109\,7\,(25)         & -0.044\,0\,(16)        & -0.003\,56\,(28)       & 681.449\,4\,(35)(1)    \\
 63     &  $^{153}$Eu & 5.1115\,(62) &   730.171\,6\,(19)(2)    & 1.167\,9\,(28)         & -0.042\,4\,(17)        & -0.003\,81\,(31)       & 731.293\,3\,(38)(2)    \\
 64     &  $^{158}$Gd & 5.1569\,(43) &   782.845\,4\,(20)(2)    & 1.227\,3\,(31)         & -0.040\,4\,(18)        & -0.003\,98\,(34)       & 784.028\,3\,(41)(2)    \\
 65     &  $^{159}$Tb & 5.06\,(15)   &   838.545\,6\,(21)(43)   & 1.287\,7\,(34)         & -0.037\,9\,(19)        & -0.004\,3\,(4)         & 839.791\,1\,(44)(43)   \\
 66     &  $^{162}$Dy & 5.207\,(17)  &   897.402\,5\,(23)(6)    & 1.348\,9\,(37)         & -0.034\,9\,(20)        & -0.004\,5\,(4)         & 898.712\,1\,(48)(6)    \\
 67     &  $^{165}$Ho & 5.202\,(31)  &   959.570\,4\,(24)(12)   & 1.410\,8\,(41)         & -0.031\,2\,(21)        & -0.004\,7\,(5)         & 960.945\,2\,(52)(12)   \\
 68     &  $^{166}$Er & 5.2516\,(31) &   1025.194\,8\,(25)(2)   & 1.473\,1\,(44)         & -0.026\,9\,(22)        & -0.005\,0\,(5)       & 1\,026.635\,9\,(56)(2)   \\
 69     &  $^{169}$Tm & 5.2256\,(35) &   1094.437\,3\,(27)(3)   & 1.535\,6\,(49)         & -0.021\,9\,(23)        & -0.005\,3\,(6)       & 1\,095.945\,7\,(60)(3)   \\
 70     &  $^{174}$Yb & 5.3108\,(60) &   1167.454\,8\,(28)(4)   & 1.598\,1\,(31)         & -0.016\,1\,(24)        & -0.005\,6\,(6)       & 1\,169.031\,3\,(49)(4)   \\
 71     &  $^{175}$Lu & 5.370\,(30)  &   1244.423\,2\,(29)(21)  & 1.660\,3\,(34)         & -0.010\,0\,(26)        & -0.005\,9\,(7)       & 1\,246.067\,7\,(52)(21)  \\
 72     &  $^{180}$Hf & 5.3470\,(32) &   1325.525\,1\,(31)(5)   & 1.721\,9\,(37)         & -0.003\,0\,(27)        & -0.006\,1\,(8)       & 1\,327.237\,9\,(56)(5)   \\
 73     &  $^{181}$Ta & 5.3507\,(34) &   1410.941\,1\,(33)(5)   & 1.782\,5\,(41)         & 0.004\,9\,(29)         & -0.006\,5\,(9)       & 1\,412.722\,0\,(60)(5)   \\
 74     &  $^{184}$W  & 5.3658\,(23) &   1500.866\,3\,(34)(6)   & 1.841\,7\,(45)         & 0.013\,8\,(30)         & -0.006\,9\,(10)      & 1\,502.715\,0\,(65)(6)   \\
 75     &  $^{187}$Re & 5.370\,(17)  &   1595.505\,7\,(36)(21)  & 1.899\,2\,(49)         & 0.024\,2\,(32)         & -0.007\,2\,(11)      & 1\,597.421\,8\,(69)(21)  \\
 76     &  $^{192}$Os & 5.4126\,(15) &   1695.066\,7\,(38)(8)   & 1.954\,4\,(53)         & 0.035\,7\,(33)         & -0.007\,5\,(12)      & 1\,697.049\,2\,(74)(8)   \\
 77     &  $^{193}$Ir & 5.40\,(11)   &   1799.781\,(4)(16)      & 2.006\,8\,(58)         & 0.048\,5\,(35)         & -0.008\,0\,(14)      & 1\,801.828\,(8)(16)    \\
 78     &  $^{196}$Pt & 5.4307\,(27) &   1909.870\,6\,(42)(11)  & 2.056\,0\,(64)         & 0.062\,6\,(37)         & -0.008\,7\,(14)      & 1\,911.980\,5\,(86)(11)  \\
 79     &  $^{197}$Au & 5.4371\,(38) &   2025.586\,5\,(44)(14)  & 2.101\,2\,(70)         & 0.078\,2\,(38)         & -0.009\,0\,(17)      & 2\,027.756\,9\,(92)(14)  \\
 80     &  $^{202}$Hg & 5.4648\,(33) &   2147.176\,7\,(46)(16)  & 2.141\,8\,(76)         & 0.095\,4\,(40)         & -0.009\,7\,(17)      & 2\,149.404\,1\,(99)(16)  \\
 81     &  $^{205}$Tl & 5.4759\,(26) &   2274.914\,3\,(48)(17)  & 2.177\,1\,(83)         & 0.114\,0\,(43)         & -0.009\,8\,(20)      & 2\,277.196\,(11)(2)    \\
 82     &  $^{208}$Pb & 5.5012\,(13) &   2409.073\,3\,(50)(19)  & 2.206\,3\,(90)         & 0.134\,5\,(45)         & -0.010\,9\,(20)      & 2\,411.403\,(11)(2)    \\
 83     &  $^{209}$Bi & 5.5211\,(26) &   2549.951\,0\,(52)(23)  & 2.228\,6\,(76)         & 0.157\,2\,(47)         & -0.010\,9\,(25)      & 2\,552.326\,(11)(2)    \\
 84     &  $^{209}$Po & 5.527\,(18)  &   2697.858\,7\,(55)(71)  & 2.242\,8\,(84)         & 0.182\,1\,(51)         & -0.011\,6\,(28)      & 2\,700.272\,(12)(7)    \\
 85     &  $^{210}$At & 5.539\,(55)  &   2853.114\,(6)(24)      & 2.248\,1\,(92)         & 0.209\,1\,(55)         & -0.012\,3\,(32)      & 2\,855.559\,(13)(24)   \\
 86     &  $^{222}$Rn & 5.691\,(20)  &   3016.003\,(6)(11)    & 2.243\,(10)          & 0.238\,(6)           & -0.012\,(3)                & 3\,018.472\,(14)(11)   \\
 87     &  $^{223}$Fr & 5.695\,(18)  &   3186.992\,(6)(11)    & 2.227\,(11)          & 0.270\,(6)           & -0.013\,(4)                & 3\,189.476\,(15)(11)   \\
 88     &  $^{226}$Ra & 5.721\,(29)  &   3366.393\,(6)(19)    & 2.199\,(12)          & 0.305\,(6)           & -0.014\,(4)                & 3\,368.883\,(16)(19)   \\
 89     &  $^{227}$Ac & 5.670\,(57)  &   3554.660\,(7)(42)    & 2.156\,(14)          & 0.343\,(6)           & -0.014\,(5)                & 3\,557.145\,(17)(42)   \\
 90     &  $^{232}$Th & 5.785\,(12)  &   3752.058\,(7)(12)    & 2.098\,(15)          & 0.384\,(7)           & -0.016\,(4)                & 3\,754.525\,(19)(12)   \\
 91     &  $^{231}$Pa & 5.700\,(57)  &   3959.276\,(7)(56)    & 2.023\,(17)          & 0.430\,(7)           & -0.016\,(6)                & 3\,961.712\,(21)(56)   \\
 92     &  $^{238}$U  & 5.8571\,(33) &   4176.439\,(8)(5)     & 1.928\,(18)          & 0.478\,(7)           & -0.016\,(6)                & 4\,178.830\,(21)(5)    \\
 \colrule\hline\\
\end{longtable*}

\end{document}